# Asymptotic Spectral Efficiency of Multi-antenna Links in Wireless Networks with Limited Tx CSI

Siddhartan Govindasamy, *Member*, Daniel W. Bliss, *Member*, and David H. Staelin, *Life Fellow, IEEE*

## Abstract

An asymptotic technique is presented for finding the spectral efficiency of multi-antenna links in wireless networks where transmitters have Channel-State-Information (CSI) corresponding to their target receiver. Transmitters are assumed to transmit independent data streams on a limited number of channel modes which limits the rank of transmit covariance matrices. This technique is applied to spatially distributed networks to derive an approximation for the asymptotic spectral efficiency in the interference-limited regime as a function of link-length, interferer density, number of antennas per receiver and transmitter, number of transmit streams and path-loss exponent. It is found that targeted-receiver CSI, which can be acquired with low overhead in duplex systems with reciprocity, can increase spectral efficiency several fold, particularly when link lengths are large, node density is high or both. Additionally, the per-link spectral efficiency is found to be a function of the ratio of node density to the number of receiver antennas, and that it can often be improved if nodes transmit using fewer streams. These results are validated for finite-sized systems by Monte-Carlo simulation and are asymptotic in the regime where the number of users and antennas per receiver approach infinity.

## I. INTRODUCTION

Multiple antenna systems are attractive for use in point-to-point and ad-hoc wireless networks since they can suppress interference, increase data rates by spatial multiplexing or coherent signal combining, and are robust to channel variations. The increased robustness available through the increased diversity and multiplexing capability of multi-antenna systems in isolated Additive White Gaussian Noise (AWGN) channels has been well studied and is described in detail in wireless communications texts such as [2]. The major contributions in this area which addressed the multiplexing capability of multi-antenna systems in AWGN channels include [3], [4], [5]. The trade-off between diversity (which increases robustness) and multiplexing (which increases data rates) was studied in [6].

A central assumption that determines the performance of Multiple-Input-Multiple-Output (MIMO) systems is the availability of Channel-State-Information (CSI) at the transmit(Tx) or receive(Rx) side of links. It is known that to achieve capacity in Additive-White-Gaussian-Noise (AWGN) channels with Rx CSI and without Tx CSI, the transmitter should send equal power, independent data streams on each antenna. With Tx CSI, capacity is achieved by parallelizing the channel between the Tx and Rx using a Singular Value Decomposition (SVD) of the channel matrix. Most works on multi-antenna systems in the literature assume that receivers have CSI, an assumption which is realistic since receivers can estimate channel parameters from signals received from transmitters.

The performance of multi-antenna systems in the presence of interference has been relatively less well studied although the ability of antenna arrays to suppress undesired signals is well known in the signal processing community (e.g. see [7]). MIMO communications systems employing interference suppression by transmitters and receivers were analyzed under different CSI assumptions by [8] and [9].

The main contributions of this article are a technique to find the asymptotic spectral efficiency of multi-antenna links in ad-hoc wireless networks with limited CSI at transmitters and arbitrary distribution of interference powers

S. Govindasamy is with the Franklin W. Olin College of Engineering. D. W. Bliss and D. H. Staelin are with the Research Laboratory of Electronics, Massachusetts Institute of Technology (MIT). D. W. Bliss is also with MIT Lincoln Laboratory. (email: siddhartan.govindasamy@olin.edu, staelin@mit.edu, bliss@ll.mit.edu). This research was supported in part by the National Science Foundation under Grant ANI-0333902. Opinions, interpretations, conclusions, and recommendations are those of the authors and are not necessarily endorsed by the United States Government. Portions of this material have been presented at the Asilomar Conference on Signals, Systems and Computers 2009, Pacific Grove CA, USA [1].

subject to a convergence criteria. In particular, we find an expression for the spectral efficiency in the interference-limited regime when interference is due to spatially distributed nodes. Transmitters are assumed to have CSI between themselves and their target receivers but to no other receiver in the network. We refer to this form of CSI as Tx-Link CSI. We assume that each transmitter is restricted to sending $M$ independent data streams on $M$ channel modes, and allow arbitrary correlation between the power allocated to each stream for any given transmitter. The covariance matrices of the signals at the antennas of each transmitting node are thus matrices of rank-$M$. Transmitters are assumed to have $K \geq M$ antennas. Limiting the number of transmit data streams in this fashion is known to increase the network spectral efficiency of wireless systems as shown analytically in [10], [11] for systems without Tx CSI, and by simulation for systems with Tx CSI in [12]. Note that in some systems, the rank of the transmit covariance matrix may be greater than the total number of independent streams, e.g. if spatial repetition coding is used. However, in this work, we assume that the number of independent data streams equals the rank of the transmit covariance matrices.

The data rates achievable in wireless networks with interfering MIMO links and no Tx CSI was studied in [10] who found that transmitters should transmit a single data stream (i.e. with a rank-1 transmit covariance matrix) from one of their antennas in the high interference regime. When interference is low, the links are essentially isolated and it is optimal for transmitters to send equal-power streams on each antenna as for AWGN channels. In [13], the authors analyzed one-to-one interfering links in the regime where the number of users goes to infinity, both in systems without Tx CSI and systems with Tx Link CSI. The authors derived upper bounds to the mean network spectral efficiency (b/s/Hz) of such systems, which were found to be constant if the number of receiver antennas per node increases linearly with the number of transmitter antennas. Neither [10] or [13] model the spatial distribution of nodes however.

The distribution of nodes in space is key to understanding large wireless networks since signal and interference strengths depend on relative node locations. Spatially distributed wireless networks with multiple-antenna links and no Tx CSI were studied in [11], which found the asymptotic spectral efficiency as the number of receiver antennas $N$ and interferers in the network $n$ tend to infinity. We found that the mean spectral efficiency was a function of the ratio of the number of receiver antennas to the product of node density and the square of the link length, implying that constant mean per-link spectral efficiency can be maintained by scaling the number of receiver antennas with node density. Recently, [14] and [15] found exact expressions for the Cumulative-Distribution-Function (CDF) of the Signal-to-Interference-Ratio (SIR) in spatially distributed networks with MMSE receivers but no CSI at the transmitters with [14] considering single-stream transmission and [15] considering multi-stream transmissions. Additionally, [16] found that it is possible to scale the network spectral efficiency per unit area linearly with node density if the number of receiver antennas scales with node density by using a partial zero-forcing receiver.

Spatially distributed systems with Tx-CSI have been studied in a number of works such as [17], [18], [19], and [20]. In [17] and [18] the receiver did not use its degrees of freedom to mitigate interference whereas [19] considered a partial zero-forcing receiver structure. An asymptotic analysis was used in [20] to analyze the spectral efficiency of MIMO links in spatially distributed networks of finite area when transmitters use single-streams with maximal-ratio-transmission (the transmit-side analogue of MRC), and MMSE receivers. In that work, the authors approximate the Signal-to-Interference-plus-Noise-Ratio (SINR) as a gamma distributed random variable and find an approximation to the CDF of the SINR as the number of receiver antennas and number of nodes in a finite-area network go to infinity.

In contrast, in addition to Tx-Link CSI and multiple transmit data streams, our work assumes optimal decoding at the receiver with interference treated as spatially-colored noise, and provides a general framework which is also applicable to systems where interference powers do not depend on node locations, which is relevant to power-controlled cellular architectures. Additionally, we assume a constant and finite density of users when applying our general technique to spatially distributed networks.

Our results concerning spatially-distributed networks characterize the spectral efficiency of multi-antenna links in ad-hoc wireless networks as a function of tangible parameters such as link length, node density, number of antennas, and path-loss exponent, and are useful for system designers to explore the trade offs between increased hardware costs of using more antennas or transmit data streams. Additionally, these results enable us to compare the spectral efficiency gains that Tx Link CSI provides.

The asymptotic techniques used here are closely related to several works in the literature such as [21] and [22]. In these works, the Signal-to-Interference-plus-Noise-Ratio (SINR) of random Code-Division-Multiple-Access (CDMA) systems, given by terms of the form $\mathbf{s}^\dagger \mathbf{R} \mathbf{s}$, are shown to converge to asymptotic limits that depend on the structure of the limiting distribution of the eigenvalues of the covariance matrix of received interference powers $\mathbf{R}$. In both [21] and [22], the vector of signatures $\mathbf{s}$ is assumed to have Independent Identically Distributed (IID) entries. Here, the SINR associated with a given stream from a transmitter has the form $\lambda \mathbf{u}^\dagger \mathbf{R} \mathbf{u}$, where $\mathbf{u}$ is a singular vector associated with a Gaussian random matrix and $\lambda$ is the square of its associated singular value. Since $\mathbf{u}$ is a unit-norm isotropic vector, its entries are not IID and hence we cannot directly apply the results of [21] and [22]. Another related work is [23] which analyzed the limiting SIR of Random CDMA systems with multiple antennas. While it allows for some correlation in the entries of the signature vector, its assumptions do not admit unit-norm isotropic vectors. Other related works include [24] who analyzed the joint asymptotic SIR distribution of multiple transmitters communicating to a single receiver in CDMA systems and [25] who analyze the asymptotic capacity in MIMO Multiple-Access and Broadcast Channels with a fixed number of users and number of antennas per user going to infinity.

Furthermore, [21], [22], and [23] analyze the limiting SINR as a function of the limiting distribution of the received interference powers, whereas we explicitly consider the distribution of transmit powers and path-losses with arbitrary correlation between the power allocated on each stream by the transmitters. While this assumption does not exclude the water-filling power allocation, we will explore practical constant-power approaches in the applications of the techniques we develop.

The remainder of the paper is organized as follows. Section II presents the network model starting with a general model of a network with one-to-one links, and followed by the spatially distributed network model. Section III contains the main results for the general network model. Section IV applies the main results to networks with a fixed path-loss from each interferer to the representative receiver, and Section V applies the general result to spatially distributed networks. Section VI contains concluding remarks and a summary of the results presented here.

The conjugate transpose of a vector $\mathbf{a}$ or matrix $\mathbf{A}$ are denoted by $\mathbf{a}^\dagger$ and $\mathbf{A}^\dagger$ respectively. The determinant is represented by $|.|$ and diagonal matrices are represented by $\mathbf{A} = diag(a_{11}, a_{22}, \cdots a_{NN})$ where $a_{ii}$ is the $ii$-th entry of the diagonal matrix $\mathbf{A}$. We use $\mathcal{CN}(0, \nu)$ to represent the circularly symmetric, complex Gaussian distribution with mean zero and variance $\nu$. The indicator function is denoted by $1_\mathcal{A}$, which equals one when the condition $\mathcal{A}$ is true and $0$ otherwise.

## II. System Model

### A. General One-to-One Network

Consider a one-to-one wireless network where there are $n + 1$ receivers and $n + 1$ transmitters where each transmitter is linked to a single receiver. Let $\mathcal{R}_i$ denote the $i$-th receiver and $\mathcal{T}_i$ denote the $i$-th transmitter. Consider a representative link between $\mathcal{R}_1$ and $\mathcal{T}_1$ where $\mathcal{T}_i$ for $i = 2, 3, \cdots n+1$ are co-channel interferers to the representative link.

The representative receiver has $N$ antennas and each transmitting node has $K \leq N$ antennas. We assume frequency-flat fading where the channel between the $j$-th antenna of transmitting node $i$ and $k$-th antenna of the representative receiver is modeled as $\sqrt{\gamma_i} g_{ijk}$, where $\gamma_i$ is the path-loss between transmitting node-$i$ and the representative receiver and $g_{ijk}$ are IID $\mathcal{CN}(0, 1)$ random variables. We make the standard assumption that nodes transmit using Gaussian code-books.

The representative receiver knows the channel co-efficients between itself and the representative transmitter, and also knows the spatial covariance matrix of the interference $\mathbf{R} = \mathbf{K}_1 \boldsymbol{\Phi}_1 \mathbf{K}_1^\dagger$, where $\mathbf{K}_1$ and $\boldsymbol{\Phi}_1$ are defined in Section III-A. Note that receivers can estimate the spatial interference covariance matrix by constructing a sample interference covariance matrix from aggregate transmissions of the interferers.

Each transmitting node knows the channel co-efficients between itself and its target receiver, but not to any other nodes. We refer to this assumption as Tx Link CSI. Note that Tx Link CSI can be obtained with low overhead in half-duplex systems with reciprocity if channels do not vary rapidly in time since transmitters can use channel estimates performed when they acted as receivers in the past, provided that the transmit and receive hardware can be accurately characterized. These channel estimates can also be performed by receivers and then fed back to the transmitters.

Additionally, we assume a thermal noise power of $N\bar{\sigma}^2$ at each antenna of the representative receiver where $\bar{\sigma}^2$ is a constant. The factor $N$ is used to ensure that the limiting SINR is finite so that we can obtain meaningful asymptotic results as $N \to \infty$ since the SINR (with a constant noise power) grows at least linearly with $N$. Equivalently, we can assume that the transmit power for each node decays inversely with $N$ as is done in works such as [21]. The asymptotic results of this paper apply in the regime as $n, N \to \infty$ with the ratio of the number of effective interferers (i.e. the total number of independent interfering streams) to receiver antennas $nM/N = c > 0$. All asymptotic results shall refer to this regime.

### B. Spatially Distributed Network

Consider a circular wireless network of radius $R$ with $n$ wireless transmitters at random IID points in the circle such that:

$$n = \rho \pi R^2. \tag{1}$$

The representative receiver $\mathcal{R}_1$ is assumed to be at the center of the circle (defined as the origin) and $\mathcal{T}_1$ is an additional transmitter at a distance $r_1$ from $\mathcal{R}_1$ as shown in Figure 1. The $n$ interferers are in links with other receivers whose locations do not impact the representative link. Let $r_i$ denote the distance between transmitting node $i$ and the origin. The path-loss $\gamma_i = G_t r_i^{-\alpha}$ with $\alpha > 2$, which is a standard model for spatially distributed networks.

For the spatially distributed network model, we shall assume a receiver noise power $N\bar{\sigma}^2$ where:

$$\bar{\sigma}^2 = \sigma^2 \left( N^{-\frac{\alpha}{2}} \right) \tag{2}$$

where $\sigma^2$ is a constant[1].

This assumption enables the asymptotic analysis of the SINR in the interference-limited regime as $N \to \infty$. Without this assumption, as $N \to \infty$, the optimal receiver will suppress interference to levels comparable to the thermal noise which results in the system no longer being interference-limited. Since we focus on the interference-limited regime for the spatially distributed network model, $\sigma^2$ is assumed to be small and does not effect the final results significantly.

## III. PARALLELIZED SYSTEM

### A. Parallelized Transmissions with Link CSI

Since transmitters do not know the channel coefficients between themselves and receivers other than their target receivers, they are not able to encode their transmissions to minimize interference they cause to unintended receivers by choice of transmit directions. Since the channels between all pairs of antennas are assumed to be Gaussian distributed, no choice of transmit directions is better than any other in terms of interference caused on unintended receivers, although the powers allocated to transmit streams can still influence the spectral efficiency of the network. Hence, it is optimal for nodes to parallelize the channels between themselves and their targets using an SVD and transmit independent data streams on each parallelized channel with some power allocation. We shall assume that all transmit nodes parallelize the channels between themselves and their respective receivers and transmit independent data streams on $M$ parallel channels where $P_{ij}$ for $i = 1, 2, \cdots n + 1$ and $j = 1, 2, \cdots, M$ denotes the power allocated to the $j$-th stream by the $i$-th transmitter. We shall refer to this as the parallelized system. Let the $P_{ij}$ be IID over $i$, i.e. the power allocations of a given transmitter are independent of other transmitters. For a given transmitter, there can be arbitrary correlation between the transmit powers it allocates to its $M$ streams. Let the Probability-Density-Function (PDF) and CDF of $P_{ij}$ for all $i$ and each $j$ be denoted by $f_j(x)$ and $F_j(x)$ respectively, and the total transmit power for each node be bounded by $P_M$, i.e. $\sum_{j=1}^{M} P_{ij} \leq P_M$ for each $i$.

Let the $N \times K$ matrix $\sqrt{\gamma_{ij}} \mathbf{H}_{ij}$ denote the channel matrix between nodes $i$ and $j$ where $\gamma_{ij}$ is the path-loss between $\mathcal{T}_i$ and $\mathcal{R}_j$, and $\mathbf{H}_{ij}$ is a matrix of IID $\mathcal{CN}(0,1)$ entries (recall that $N$ is the number of receiver antennas and $K$ is the number of antennas per transmitter). The spectral efficiency of the representative link is given by (e.g. see [26]):

---

[1]Note that the asymptotic SINR we derive for the spatially distributed network is normalized by $N^{\alpha/2}$ which is accomplished by scaling the interference and noise powers by $N^{\alpha/2}$. This scaling of the noise requires the noise power to be given by (2) so that the scaled thermal noise value equals $N\sigma^2$ which is the form of the thermal noise power for the network model of Section II-A

$$C_1 = \log_2 \left| \mathbf{I} + \gamma_1 \mathbf{H}_{11} \mathbf{T}_1 \mathbf{H}_{11}^\dagger \left( N\bar{\sigma}^2 \mathbf{I} + \sum_{j=2}^{n+1} \gamma_j \mathbf{H}_{j1} \mathbf{T}_j \mathbf{H}_{j1}^\dagger \right)^{-1} \right| \tag{3}$$

where $\mathbf{T}_j$ is the transmit covariance matrix of node-$j$, i.e., it is the covariance matrix of the signals sent on the transmit antennas of node-$j$.

Performing an SVD on $\mathbf{H}_{ij}$ yields

$$\mathbf{H}_{ij} = \mathbf{U}_{ij} \mathbf{\Sigma}_{ij} \mathbf{V}_{ij}^\dagger. \tag{4}$$

where $\mathbf{U}_{ij}$ and $\mathbf{V}_{ij}$ are unitary matrices and $\mathbf{\Sigma}_{ij}$ is a matrix containing the squared-singular values of $\mathbf{H}_{ij}$. Let $\lambda_{ij}$ denote the square of the $j$-th largest singular value of $\mathbf{H}_{i1}$. The spectral efficiency of the representative link can then be bounded according to the following Lemma proved in Appendix A

*Lemma 1:* The spectral efficiency of the representative link can be bounded from above as follows:

$$C_1 \leq \sum_{j=1}^{M} \log_2 \left( 1 + \gamma_1 P_{1j} \lambda_{1j} \mathbf{u}_{1j}^\dagger \left( N\bar{\sigma}^2 \mathbf{I} + \mathbf{K}_1 \mathbf{\Phi}_1 \mathbf{K}_1^\dagger \right)^{-1} \mathbf{u}_{1j} \right) \tag{5}$$

and below as follows:

$$C_1 \geq \sum_{j=1}^{M} \log_2 \left( 1 + \gamma_1 P_{1j} \lambda_{1j} \hat{\mathbf{u}}_j^\dagger \left( N\bar{\sigma}^2 \mathbf{I} + \hat{\mathbf{K}}_j \mathbf{\Phi}_1 \hat{\mathbf{K}}_j^\dagger \right)^{-1} \hat{\mathbf{u}}_j \right) \tag{6}$$

where

$$\mathbf{\Phi}_1 = diag(\gamma_2 P_{21}, \cdots, \gamma_2 P_{2M}, \gamma_3 P_{31}, \cdots, \gamma_3 P_{3M}, \cdots, \gamma_{n+1} P_{(n+1)1}, \cdots, \gamma_{n+1} P_{(n+1)M}) \tag{7}$$

The entries of the $N \times nM$ matrix $\mathbf{K}_1$ are IID $\mathcal{CN}(0,1)$ random variables and are defined by (52) in Appendix A, $\mathbf{u}_{1j}$ are $N \times 1$, unit-norm isotropic random vectors that are mutually orthogonal, and recall that $\gamma_i$ is the path-loss between the $i$-th transmitter and the representative receiver. For the lower bound, $\hat{\mathbf{K}}_j$ are $(N - M + 1) \times nM$ matrices with IID $\mathcal{CN}(0,1)$ entries, and $\hat{\mathbf{u}}_j$ are $(N - M + 1) \times 1$, unit-norm, isotropic random vectors.

Note that the upper bound corresponds to the case where the $M$ streams from the representative transmitter do not interfere with each other. The lower bound is achieved when the receiver uses $M - 1$ of its $N$ degrees of freedom to completely null the interference from the $M - 1$ other streams of the representative transmitter.

### B. Asymptotic Spectral Efficiency of the Parallelized System

The asymptotic spectral efficiency of the parallelized system described in the previous sub-section is characterized by the following theorem. Note that Theorem 1 is presented for the general assumptions in Section II-A and does not rely on the spatial distribution of nodes or the path loss model described in Section II-B.

*Theorem 1:* Let $\Psi(\tau)$ denote the CDF of the path-losses from the interferers to the representative receiver $\gamma_2, \gamma_3, \cdots, \gamma_{n+1}$. In the limit as $n, N \to \infty$ with the ratio of interferers to receiver antennas $nM/N = c$, the spectral efficiency of link-1 converges with probability 1 to

$$C_1^* = \sum_{j=1}^{M} \log_2(1 + \lambda_j^* P_{1j} \gamma_1 \beta) \tag{8}$$

where $\beta$ is a unique, non-negative solution to the equation:

$$-\bar{\sigma}^2 \beta + 1 = \beta c \int_0^\infty \frac{x\, dH(x)}{1 + x\beta} \tag{9}$$

and $H(x)$ is the limit of the empirical distribution function of the received interference powers given by:

$$H(x) = \frac{1}{M} \sum_{j=1}^{M} \int f_j(\tau) \Psi(x/\tau) \, d\tau \tag{10}$$

and $\lambda_j^*$ is the limiting value of the $j$-th largest eigenvalue of the Wishart matrix $\frac{1}{N} \mathbf{G}\mathbf{G}^\dagger$ where $\mathbf{G}$ is an $N \times K$ matrix with IID $\mathcal{CN}(0,1)$ entries. In particular if $N, K \to \infty$ such that $K/N = d$ with $0 < d \leq 1$, then:

$$\lambda_1^* = \lambda_2^* = \cdots = \lambda_M^* = (1 + \sqrt{d})^2 \tag{11}$$

and if $K$ is a finite constant,

$$\lambda_1^* = \lambda_2^* = \cdots = \lambda_M^* = 1. \tag{12}$$

*Proof:* Consider the upper bound for the spectral efficiency of the representative link given in Lemma 1, and Lemma 2 with $m = 0$. The upper bound from Lemma 1 can now be written as:

$$C_1 \leq \sum_{j=1}^{M} \log_2(1 + \text{SINR}_j) \tag{13}$$

Since 1) $\text{SINR}_j$ converges with probability 1 to $\lambda_j^* P_{1j} \gamma_1 \beta$ from Lemma 2, 2) the log function is continuous and 3) the sum of terms that converge with probability 1 converges with probability 1 to the sum of the limits (e.g. see [27]) the RHS of (13) converges with probability 1 to

$$\sum_{j=1}^{M} \log_2(1 + \lambda_j^* P_{1j} \gamma_1 \beta). \tag{14}$$

Now, let $m = M - 1$. In this case, the lower bound can be written as:

$$C_1 \geq \sum_{j=1}^{M} \log_2(1 + \text{SINR}_j) \to \sum_{j=1}^{M} \log_2(1 + \lambda_j^* P_{1j} \gamma_1 \beta) \tag{15}$$

The convergence is due to the fact that $\text{SINR}_j$ converges with probability 1 to $\lambda_j^* P_{1j} \gamma_1 \beta$ from Lemma 2, independent of $m$.

Since the upper and lower bounds converge to the same value, we conclude that $C_1$ converges with probability 1 to $\sum_{j=1}^{M} \log_2(1 + \lambda_j^* P_{1j} \gamma_1 \beta)$. ∎

$\beta$ can be interpreted as the limiting Rx array gain SINR as it is the limit of the SINR when the effects of transmit power, transmit beamforming, and path-loss are not taken into account.

*Lemma 2:* Define $\text{SINR}_j$, which can represent either the lower or upper bound of the SINR of the $j$-th stream from the representative transmitter, as follows:

$$\text{SINR}_j = \gamma_1 P_{1j} \lambda_{1j} \mathbf{w}_j^\dagger \left( N \bar{\sigma}^2 \mathbf{I} + \mathbf{K} \mathbf{\Phi}_1 \mathbf{K}^\dagger \right)^{-1} \mathbf{w}_j. \tag{16}$$

where $\mathbf{K}$ is an $(N - m) \times nM$ random matrix with IID $\mathcal{CN}(0,1)$ entries with $m$ a non-negative, finite integer, and $\mathbf{w}$ is an $(N - m) \times 1$ isotropic random vector with unit norm.

Under the assumptions of Theorem 1, $\text{SINR}_j$ converges with probability 1 to an asymptotic limit given by:

$$\text{SINR}_j \to \gamma_1 P_{1j} \lambda_j^* \beta \tag{17}$$

where the limiting Rx array gain SINR $\beta$, is a unique, non-negative solution for $\beta$ in (9) and $\lambda_j^*$ are defined in Theorem 1. Note that the limit of $\text{SINR}_j$ does not depend on $m$.

*Proof:* The proof uses the main results of [23] and is presented in Appendix B.

Additionally, we note that the convergence of $\lambda_1, \lambda_2, \cdots, \lambda_M$ to either 1 or $(1 + \sqrt{d})^2$ is slow and is not a good approximation for moderate values of $N$ and $K$. Recall that $d = K/N$ is the ratio of the number of antenna at the transmitters to those at the receivers. Instead, we approximate the $j$-th largest eigenvalue by the limiting distribution of the eigenvalues evaluated at a fraction $N - j$ of the range between its minimum and maximum values. This yields the following approximation:

$$\lambda_{1j}^* \approx F_d^{-1}((N - j + 1)/N).$$ (18)

where $F_d^{-1}(x)$ is the functional inverse of $F_d(x)$ which is the limiting empirical distribution function (e.d.f.) of the eigenvalues of a Wishart matrix $\frac{1}{N}\mathbf{G}\mathbf{G}^\dagger$. $\mathbf{G}$ here is an $N \times K$ matrix of IID $\mathcal{CN}(0,1)$ entries. $F_d(x)$ is given for a general $d$ in Appendix F. For $d = 1$, i.e. each transmitter and receiver have the same number of antennas,

$$F_d(x) = \begin{cases} \frac{\pi + \sqrt{4x - x^2} + 2\arcsin\left(\frac{x}{2} - 1\right)}{2\pi} & \text{if } 0 \le x < 4 \\ 1 & \text{otherwise} \end{cases}$$ (19)

Equation (18) can be found by evaluating $F_d^{-1}(x)$ numerically.

## IV. Application to Constant Path-loss Systems

### A. Asymptotic Spectral Efficiency

To test the form of the spectral efficiency described in Theorem 1, we consider two different models for the transmit power. In the first model the transmit power used on all $M$ streams by the interferers are constants denoted by $P$, which we call the equal power model. The second model has two classes of nodes and is called the two-class model. The first class transmits with power $P_1$ on all streams and the second transmits with power $P_2$ from a single stream. The interferers are assigned to classes independently and randomly where the probability that a given node is assigned to class one equals $q$. The two-class model is useful in the context of mixed systems where some fraction of transmitters have multiple antennas and the remainder have single antennas. Additionally, the power allocated by each user to their transmit streams are correlated, which illustrates the applicability of Theorem 1 in systems with correlated transmit powers.

We assume that the path-losses from all interferers to the representative node equal a constant $\gamma$ so that the CDF of path-losses $\Psi(x)$ can expressed in terms of a "step" function with a step at $\gamma$, and (9) which is the implicit equation for the limiting Rx array gain SINR $\beta$ becomes:

$$-\bar{\sigma}^2\beta + 1 = \beta c \int_0^\infty \frac{1}{M}\sum_{j=1}^M \frac{x f_j(x/\gamma)}{1 + x\beta}\, dx.$$ (20)

For the equal power model, $f_i(x) = \delta(x - P)$ and (20) becomes

$$-\bar{\sigma}^2\beta + 1 = \frac{\beta c P \gamma}{1 + P\gamma\beta}$$ (21)

Applying the quadratic formula and selecting the positive term:

$$\beta_{ep} = \frac{1 - c}{2\bar{\sigma}^2} - \frac{1}{2P\gamma} + \sqrt{\frac{(1-c)^2}{4\bar{\sigma}^4} + \frac{1+c}{2P\gamma\bar{\sigma}^2} + \frac{1}{4P^2\gamma^2}}$$ (22)

where we use the notation $\beta_{ep}$ to denote the limiting Rx array gain SINR for the equal-power model. Thus, from Theorem 1, the limiting spectral efficiency on the representative link (link 1) is:

$$C_{1ep} = \sum_{j=1}^M \log_2(1 + \lambda_j^* P \gamma_1 \beta_{ep})$$ (23)

For the two-class model, the marginal PDFs of the transmit powers are:

$$f_j(x) = \begin{cases} q\delta(x - P_1) + (1 - q)\delta(x - P_2) & \text{for } j = 1 \\ q\delta(x - P_1) & \text{for } j = 2, \cdots, M \end{cases}$$ (24)

and (9) becomes

$$-\bar{\sigma}^2\beta + 1 = q\frac{\beta c P_1 \gamma}{1 + P_1\gamma\beta} + (1 - q)\frac{\beta c P_2 \gamma}{M(1 + P_2\gamma\beta)}$$ (25)

The exact solution for the limiting Rx array gain SINR for the two-class model is found by solving (25). The solution denoted by $\beta_{2c}$, is given by (89) in Appendix G. Hence, the asymptotic spectral efficiency for the two-class model is:

$$C_{12c} = \sum_{j=1}^{M} \log_2(1 + \lambda_j^* P_{1j} \gamma_1 \beta_{2c}).$$ (26)

### B. Monte-Carlo Simulations

We verified the expressions for the asymptotic spectral efficiency under the constant transmit power and two-class models, given by (23) and (26) respectively, with $\lambda_{1j}^*$ approximated by (18). In both cases, we assumed $\bar{\sigma}^2 = 1 \times 10^{-13}$ $W$ with the thermal noise power equaling $N\bar{\sigma}^2$.

We simulated systems with the ratio of interferers to receiver antennas $n/N = 1$ and $n/N = 4$ with a common path loss to the representative receiver of $\gamma = -125$ dB. The representative transmitter had a path loss of $-100$ dB to the representative receiver. Each experiment was repeated 1000 times.

For the constant transmit power model, each transmitting node transmitted $P = \frac{1}{M}$ W on each of the $M$ data streams. For the two-class model, each interferer was class one with probability $q = 0.5$ where class-one interferers transmitted with power $P_1 = 0.5$ W on each of $M$ streams and class-two interferers transmitted $P_2 = 1$ W on a single stream. The representative transmitter was always designated as a class-one transmitter. For both models we simulated $M = 1, 2, 4$ and 8 streams per transmitting node with equal numbers of antennas at all transmitters and at the representative receiver, i.e. $N = K$.

Figures 2 and 3 show results of the simulations for the constant power model with $n/N = 1$ and $n/N = 4$, respectively, and the asymptotic spectral efficiency predicted by (23). The points represent a random sampling of 100 trials from the 1000 trials of the simulation for each $N$. The standard deviation of the spectral efficiency from 1000 trials is plotted using dashed lines. The convergence of the spectral efficiency is evident from the figure since the points representing different trials of the simulation converge with increasing $N$. Additionally, note that the standard deviation decays with $N$ which indicates convergence in the mean-square sense. For $N \geq 14$ antennas and 1000 trials, the largest deviation from the asymptotic prediction is less than 15% for $n/N = 1$ and $n/N = 4$. For $N \geq 25$, the largest deviation falls below 10% in both cases.

Figure 4 show results of the simulations for the two-class model with $n = 128$ interferers and the asymptotic spectral efficiency predicted by (26). For $N \geq 14$ antennas and from 1000 trials, the largest deviation from the asymptotic prediction is less less than 15% for $n/N = 4$, which is similar to the constant power model, and remains below 10% for $N \geq 28$.

## V. Application to Spatially Distributed Networks

### A. Asymptotic Spectral Efficiency

We now apply the results of Section III-B to the spatially distributed network model of Section II-B. In this case it is known from [11] that as the number of receiver antennas $N \to \infty$, the SINR in the interference-limited regime for systems without Tx CSI grows as $N^{\alpha/2}$ where $\alpha$ is the path-loss exponent. To avoid singularities, we define a normalized SINR for the $j$-th data stream from the representative receiver as follows:

$$\eta_{jN} = N^{-\alpha/2}\overline{\mathrm{SINR}}_j.$$ (27)

where $\overline{\mathrm{SINR}}_j$ is the SINR associated with the $j$-th data stream from the representative transmitter.

This normalization is accomplished by scaling the path-loss of the interferers and the thermal noise by $N^{\alpha/2}$. The normalized SINR is simply the SINR of this new system with the scaled interference and noise powers. The limiting value of the normalized SINR is given by the following theorem which applies Lemma 2 to the scaled interference-power model.

*Theorem 2:* As the number of interferers $n$, receiver antennas $N$, and network radius $R \to \infty$ with $nM/N = c$ and $n = \pi\rho R^2$, the normalized SINR for stream-$i$ $\eta_{iN}$ approaches an asymptotic limit with probability 1 as follows:

$$\eta_{iN} \to \eta_i = P_{1i}\lambda_i^* G_t r_1^{-\alpha}\beta$$ (28)

where the limiting Rx array-gain SINR $\beta$ satisfies the following equation:

$$\frac{2\pi^2\rho\left(G_t\beta\right)^{\frac{2}{\alpha}}}{\alpha}\left(\sum_{j=1}^{M}E\left[P_j^{\frac{2}{\alpha}}\right]\csc\left(\frac{2\pi}{\alpha}\right)\right)$$

$$-\frac{2\pi\rho\beta}{\alpha}\left(\int_0^\infty\frac{\tau^{-\frac{2}{\alpha}}}{1+\tau\beta}\,d\tau\right)\left(\sum_{j=1}^{M}\int_{\frac{\tau}{G_t b}}^\infty f_j(x)x^{\frac{2}{\alpha}}dx\right)+\beta\sigma^2=1 \qquad (29)$$

where $b=\left(\frac{\pi\rho N}{n}\right)^{\alpha/2}$ and $E[P_j^{2/\alpha}]$ is the expected value of the transmit power allocated by the interferers to their $j$-th strongest stream, raised to the power $\frac{2}{\alpha}$.

*Proof:* The proof is presented in Appendix D.

In general, (29) has to be solved numerically. Moreover, the relationship between parameters such as the number of receiver antennas, interferer density, $\alpha$ and the SINR is not clear from (29). However if we assume that $\beta$ is a continuous function of $c$, a few approximations can be made to yield additional insight into how the various factors contribute to the limiting SINR, starting with the following Lemma proved in Appendix H:

*Lemma 3:* As $b=\left(\frac{\pi\rho N}{n}\right)^{\alpha/2}\to 0$, if the total transmit power per node is bounded from above by $P_M>0$, then

$$\int_0^\infty\frac{\tau^{-\frac{2}{\alpha}}}{1+\tau\beta}\,d\tau\int_{\frac{\tau}{G_t b}}^\infty f_j(x)x^{\frac{2}{\alpha}}dx\to 0. \qquad (30)$$

for $j=1,2,\cdots M$.

From Lemma 3, we note that if $n/N$ is very large i.e. the number of nodes in the network is much larger than the number of antennas per receiver, the transmit power limit per node implies that the second term on the LHS of (29) is small. Furthermore, we assume that the thermal noise power is small, which implies that the third term on the LHS of (29) is small. Using these approximations we approximate the limiting Rx array gain SINR $\beta$ as follows:

$$\frac{2\pi^2\rho\left(G_t\beta\right)^{\frac{2}{\alpha}}}{\alpha}\sum_{j=1}^{M}E\left[P_j^{\frac{2}{\alpha}}\right]\csc\left(\frac{2\pi}{\alpha}\right)\approx 1$$

$$\beta\approx\frac{1}{G_t}\left[\frac{\alpha}{2\pi^2\rho\sum_{j=1}^{M}E\left[P_j^{\frac{2}{\alpha}}\right]}\sin\left(\frac{2\pi}{\alpha}\right)\right]^{\alpha/2} \qquad (31)$$

Substituting (31) into (28) yields a normalized SINR of:

$$\eta_i\approx\lambda_{1i}^*P_{1i}\left[\frac{\alpha}{2\pi^2\rho r_1^2\sum_{j=1}^{M}E\left[P_j^{\frac{2}{\alpha}}\right]}\sin\left(\frac{2\pi}{\alpha}\right)\right]^{\alpha/2}. \qquad (32)$$

Since all transmissions use Gaussian code-books, we use the Shannon formula for the spectral efficiency for a given link. Writing $G_\alpha=\left(\frac{\alpha}{2\pi}\sin\left(\frac{2\pi}{\alpha}\right)\right)^{\alpha/2}$ and summing the contribution from $M$ streams yields the following approximation for the spectral efficiency (with the SINR normalization) of link 1 when $N$ is large:

$$\bar{C}_1\approx\sum_{i=1}^{M}\log_2\left(1+\lambda_{1i}^*P_{1i}G_\alpha\left[\frac{1}{\pi\rho r_1^2\sum_{j=1}^{M}E\left[P_j^{\frac{2}{\alpha}}\right]}\right]^{\frac{\alpha}{2}}\right) \qquad (33)$$

where $r_1$ is the length of the representative link and $P_{1i}$ is the transmit power allocated by the representative transmitter to its $i$-th stream.

Removing the normalization by $N^{\alpha/2}$ we approximate the spectral efficiency of link-1 for large $N$ as:

$$C_1 \approx \sum_{i=1}^{M} \log_2 \left( 1 + \lambda_{1i}^* P_{1i} G_\alpha \left[ \frac{N}{\pi \rho r_1^2 \sum_{j=1}^{M} E \left[ P_j^{\frac{2}{\alpha}} \right]} \right]^{\frac{\alpha}{2}} \right). \tag{34}$$

Note that the RHS of (34) grows with $N$ and hence does not converge. However, the asymptotic spectral efficiency of (34) is a good estimate of the mean spectral efficiency for large $N$ since the deviation of the mean spectral efficiency from the asymptotic spectral efficiency is small, as shown in Appendix E. Hence,

$$E[C_1] \approx \sum_{i=1}^{M} \log_2 \left( 1 + \lambda_{1i}^* P_{1i} G_\alpha \left[ \frac{N}{\pi \rho r_1^2 \sum_{j=1}^{M} E \left[ P_j^{\frac{2}{\alpha}} \right]} \right]^{\frac{\alpha}{2}} \right). \tag{35}$$

For the equal power model, (35) becomes:

$$E\left[ C_{1ep} \right] \approx \sum_{i=1}^{M} \log_2 \left( 1 + \lambda_{1i}^* G_\alpha \left[ \frac{N}{M \pi \rho r_1^2} \right]^{\frac{\alpha}{2}} \right). \tag{36}$$

For the two-class model with Link 1 assigned to class-one, (33) becomes:

$$\bar{C}_{12c} \approx \sum_{i=1}^{M} \log_2 \left( 1 + \lambda_{1i}^* P_1 G_\alpha \left[ \frac{1}{\pi \rho r_1^2 (q M P_1^{\frac{2}{\alpha}} + (1-q) P_2^{\frac{2}{\alpha}})} \right]^{\alpha/2} \right) \tag{37}$$

and (35) becomes:

$$E\left[ C_{12c} \right] \approx \sum_{i=1}^{M} \log_2 \left( 1 + \lambda_{1i}^* P_1 G_\alpha \left[ \frac{N}{\pi \rho r_1^2 (q M P_1^{\frac{2}{\alpha}} + (1-q) P_2^{\frac{2}{\alpha}})} \right]^{\alpha/2} \right). \tag{38}$$

### B. Monte-Carlo Simulations

We simulated spatially distributed systems to validate (36) and (38). We placed 1000 interferers at random locations within a circle of radius selected such that the density of nodes in the network was $10^{-3}$ nodes per unit area and the link-length $r_1$ was such that $\pi \rho r_1^2 = 1$. The path-loss exponent $\alpha$ was set to 3 or 4 and the thermal noise level was constant at $1 \times 10^{-13}$ W. Note that the specific value of the thermal noise power does not play a significant role in the interference-limited systems we simulated. In each case the number of antennas at the representative receiver $N$ and the transmitting nodes $K$ were equal, and for each $N$ the experiment was repeated 1000 times.

For the constant-transmit-power model, each transmitting node transmitted $P = \frac{1}{M}$ W on each of $M$ data streams. For the two-class model, each interferer was class one with probability $q = 0.5$, where class-one interferers transmitted with $P_1 = 0.5$ W on each of $M$ streams, and class-two interferers transmitted $P_2 = 1$ W on a single stream. The representative transmitter was always designated as a class-one transmitter. For both models, we simulated $M = 1, 2, 4$ and 8 streams per transmitting node, with equal numbers of antennas at all transmitters and at the representative receiver.

Figures 5 and 6 show results of simulations of spatially-distributed systems with constant transmit powers, $\alpha = 4$, and SINRs normalized by scaling the interference powers by $N^{\alpha/2}$ for 1 and 4 streams per transmitter, respectively. The solid lines represent (33) and the dashed lines represent the standard deviation of the simulated results. For clarity, only a random sampling of 100 of the 1000 trials are plotted. Figure 7 depicts the simulated spectral efficiencies with normalized SINRs for the two-class model and 4 streams per transmitter, with the solid line representing (37) with $\alpha = 4$. In all cases presented here, the spectral efficiencies clearly converge on the asymptotic limit as indicated by the distribution of points and the reduction in standard deviation [2].

---

[2]Note that the diminishing standard deviation does not in general imply convergence with probability 1 but it does imply convergence in probability.

The spectral efficiencies with un-normalized SINR given by (34) and (38) do not converge as the spectral efficiency increases with $N$. However, the asymptotic spectral efficiency is a good approximation for the mean spectral efficiency for large $N$ as shown in Appendix E.

Figures 8 and 9 respectively show the simulated mean spectral efficiencies for the constant power model and 2-class model with 1, 2, 4, and 8 streams per transmitting node and $\alpha = 4$. Equations (34) and (38) are also plotted in the figures.

With constant transmit powers per stream of $P/M$, Figure 8 indicates that for all simulated cases, the asymptotic expression is within 10 % of the simulated mean spectral efficiencies when there are greater than 10 antennas. For the two-class model the simulated mean spectral efficiency is within 10 % of the asymptotic expression when $N \geq 9$ as shown in Figure 9.

Figure 10 illustrates the simulated and asymptotic mean spectral efficiency with $\alpha = 3$ and the constant transmit power model. For reference, the asymptotic spectral efficiency for $\alpha = 4$ is plotted using dashed lines. The mean spectral efficiency converges to the asymptotic value in a fashion similar to that for $\alpha = 4$, although the mean spectral efficiency is consistently lower for $\alpha = 3$ compared to $\alpha = 4$, consistent with (34).

### C. Comparison to Systems without Transmit CSI

When transmitting nodes do not have CSI and transmit with equal power on each stream, the asymptotic mean spectral efficiency is known to be [11]:

$$E[C_{NC}] \approx M \log_2 \left( 1 + G_\alpha \left[ \frac{N}{AM} \right]^{\frac{\alpha}{2}} \right). \tag{39}$$

where $A = \pi \rho r_1^2$. This quantity can be interpreted as the average number of interferers closer to a given receiver than its target transmitter and was defined as the link *rank* in [11].

To compare the spectral efficiency with and without Tx Link CSI, we plotted the ratio of the asymptotic mean spectral efficiency with and without Tx CSI. Figure 11 shows the ratio of the mean spectral efficiency with Tx Link CSI to the mean spectral efficiency without Tx CSI (expressed as a percentage) vs. link-rank $A$ for $N = K = 8$ (dashed lines) and $N = K = 12$ (solid lines) antennas at transmitters and receivers. Note that the gain with Tx-Link CSI is highly dependent on $A$. For instance, for rank-6 links and two transmit streams, Tx-Link CSI doubles the spectral efficiency for $N = 12$. Also note that the increase in mean spectral efficiency can be greater than three-fold for high-rank links.

The increase in spectral efficiency with Tx CSI is greater for large $A$ because the SINR tends to be lower for for large $A$ and so the SINR gain provided by the Tx CSI makes a bigger difference inside the log function in the expression for the spectral efficiency. The SINR increase in Figure 11 flattens for large $A$ because at low SINR the spectral efficiency of a given stream approaches the SINR because $\log(1 + x) \approx x$ for $x \ll 1$. In this regime, the ratio of spectral efficiencies with and without Tx CSI does not change significantly with increasing $A$. The increase in spectral efficiency with Tx CSI is lower for a large number of streams due to the fact that the weaker streams have SINRs that are close to the SINRs obtained without Tx CSI.

Figure 11 thus indicates that a significant (but not an orders of magnitude) increase in spectral efficiency is possible with Tx Link CSI that can be acquired with low overhead in duplex systems provided that the transmit and receive hardware paths can be accurately characterized.

## VI. Summary and Conclusions

A technique is presented for computing the asymptotic spectral efficiency of multi-antenna links in ad-hoc wireless networks where transmitters have CSI corresponding to their desired receivers. The transmitters are restricted to using $M$ channel modes (which limits the rank of the transmit covariance matrices to $M$), and the results are asymptotic in the regime where the numbers of receiver antennas and interferers go to infinity with a constant ratio. The asymptotic predictions are supported by numerical simulations. The simulations indicate that the asymptotic expressions are good estimates for the spectral efficiency even when the number of receiver antennas is moderately large which is useful since it is possible to place approximately 20 or more antenna elements on a standard laptop computer with wave-length separation at a nominal carrier frequency of 6GHz.

In spatially distributed networks the asymptotic spectral efficiency, which approximates the mean spectral efficiency is found to be dependent on the ratio of interferer density to the number of receiver antennas, as given in (34). Thus, as is the case for systems without Tx CSI [11], it is possible to maintain a constant mean spectral efficiency per link if the number of antennas per receiver is increased linearly with interferer density.

Additionally, we found that the spectral efficiency in the network can be increased in certain cases if each transmitter transmits fewer data streams unlike MIMO links in AWGN channels where to maximize capacity, nodes should transmit data on all their channel modes with a water-filling power allocation. Figure 8 illustrates this potential advantage of transmitting fewer streams when the number of antennas is relatively small; a similar observation was made in [11] for systems without Tx CSI.

Compared to systems without Tx CSI, we find that the asymptotic spectral efficiency with Tx-Link CSI can be several times larger, where the benefit of using Tx-Link CSI increases for longer links, denser networks, or both, as illustrated by Figure 11. For instance, with four transmit streams and 12 antennas at transmitters and receivers, Tx CSI can double the spectral efficiency when the link-rank $A = 3$, where $A = \pi \rho r_1^2$, which captures link length and interferer distribution. Since Tx-Link CSI can be estimated in duplex systems with reciprocity without a significant increase in overhead, Tx Link CSI can be useful, particularly in dense networks with long links.

## VII. Acknowledgements

We thank N. Jindal for pointing out references [14] and [15] and helpful comments.

## Appendix

### A. Bounds on the Spectral Efficiency of Parallelized System

The spectral efficiency of link $i$ between $\mathcal{R}_i$ and $\mathcal{T}_i$ is given by (see e.g. [26]):

$$C_i = \log_2 \left| \mathbf{I} + \gamma_{ii} \mathbf{H}_{ii} \mathbf{T}_i \mathbf{H}_{ii}^\dagger \left( N \bar{\sigma}^2 \mathbf{I} + \sum_{j=1, j \neq i}^{n+1} \gamma_{ij} \mathbf{H}_{ji} \mathbf{T}_j \mathbf{H}_{ji}^\dagger \right)^{-1} \right| \tag{40}$$

where $\gamma_{ij}$ is the path loss between transmitter $\mathcal{T}_i$ and $\mathcal{R}_j$, and $\mathbf{T}_j$ is the transmit covariance matrix of $\mathcal{T}_j$, i.e., it is the covariance matrix of the signals sent on the transmit antennas of node $j$. Recall that $\sqrt{\gamma_{ij}} \mathbf{H}_{ij}$ is the $N \times K$ matrix of channel coefficients between the antennas of $\mathcal{T}_i$ and $\mathcal{R}_j$.

Since transmitters do not know the channels between themselves and unintended targets, they cannot choose their transmit covariance matrices so as to transmit in spatial directions that reduce interference to undesired receivers. Hence, they should transmit in spatial directions that maximize the data rate on their individual links. Note that their choice of transmit powers to allocate to their streams can still influence the spectral efficiency of other links.

It is known that without knowledge of the quantity in the parenthesis in (40), to maximize the RHS of (40) the $i$-th transmitter should use the following transmit covariance matrix:

$$\mathbf{T}_i = \mathbf{V}_i \mathbf{P}_i \mathbf{V}_i^\dagger \tag{41}$$

with

$$\mathbf{P}_i = \text{diag} \left( P_{i1}, P_{i2}, \cdots, P_{iM}, 0, \cdots \right) \tag{42}$$

where $P_{jk}$ is the power allocated to the $k$-th stream by the $j$-th transmitter. $\mathbf{V}_i$ is a unitary matrix defined by taking the SVD of $\mathbf{H}_{ii}$ such that $\mathbf{H}_{ii} = \mathbf{U}_i \mathbf{\Sigma}_i \mathbf{V}_i^\dagger$. $\mathbf{\Sigma}_i$ here is a diagonal matrix of the singular values of $\mathbf{H}_i$ with $\lambda_{ij}$ equal to the square of the $j$-th largest singular value, and $\mathbf{U}_i$ is another unitary matrix.

Assuming that all transmitters use the same strategy, substituting (41) into (40) yields:

$$C_i = \log_2 \left| \mathbf{I} + \gamma_{ii} \mathbf{H}_{ii} \mathbf{V}_i \mathbf{P}_i \mathbf{V}_i^\dagger \mathbf{H}_{ii}^\dagger \left( N \bar{\sigma}^2 \mathbf{I} + \sum_{j=1, j \neq i}^{n+1} \gamma_{ij} \mathbf{H}_{ji} \mathbf{V}_j \mathbf{P}_j \mathbf{V}_j^\dagger \mathbf{H}_{ji}^\dagger \right)^{-1} \right|. \tag{43}$$

Note that random matrices with Gaussian distributed entries maintain their statistical properties when multiplied by unitary matrices. Thus, we can write (43) as

$$C_i = \log_2 \left| \mathbf{I} + \gamma_{ii} \mathbf{H}_{ii} \mathbf{V}_i \mathbf{P}_i \mathbf{V}_i^\dagger \mathbf{H}_{ii}^\dagger \left( N\bar{\sigma}^2 \mathbf{I} + \sum_{j=1, j\neq i}^{n+1} \gamma_{ij} \tilde{\mathbf{H}}_{ji} \mathbf{P}_j \tilde{\mathbf{H}}_{ji}^\dagger \right)^{-1} \right| \tag{44}$$

where $\tilde{\mathbf{H}}_{ij}$ are distributed identically to $\mathbf{H}_{ij}$.

Substituting $\mathbf{H}_{ii} = \mathbf{U}_i \mathbf{\Sigma}_i \mathbf{V}_i^\dagger$ and noting that $\mathbf{V}_i \mathbf{V}_i^\dagger = \mathbf{I}$:

$$
\begin{aligned}
C_i &= \log_2 \left| \mathbf{I} + \gamma_{ii} \mathbf{U}_i \mathbf{\Sigma}_i \mathbf{P}_i \mathbf{\Sigma}_i^\dagger \mathbf{U}_i^\dagger \left( N\bar{\sigma}^2 \mathbf{I} + \sum_{j=1, j\neq i}^{n+1} \gamma_{ij} \tilde{\mathbf{H}}_{ji} \mathbf{P}_j \tilde{\mathbf{H}}_{ji}^\dagger \right)^{-1} \right| \\
&= \log_2 \left| \mathbf{I} + \gamma_{ii} \mathbf{\Sigma}_i \mathbf{P}_i \mathbf{\Sigma}_i^\dagger \mathbf{U}_i^\dagger \left( N\bar{\sigma}^2 \mathbf{I} + \sum_{j=1, j\neq i}^{n+1} \gamma_{ij} \tilde{\mathbf{H}}_{ji} \mathbf{P}_j \tilde{\mathbf{H}}_{ji}^\dagger \right)^{-1} \mathbf{U}_i \right|
\end{aligned}
\tag{45}
$$

The steps from (40) to (45) are standard and can be found in [26].

*1) Upper Bound on Spectral Efficiency:* Consider the spectral efficiency of the representative link, link 1. For notational convenience, define:

$$\mathbf{Q} = \mathbf{U}_1^\dagger \left( N\bar{\sigma}^2 \mathbf{I} + \sum_{j=2}^{n+1} \gamma_{1j} \tilde{\mathbf{H}}_{j1} \mathbf{P}_j \tilde{\mathbf{H}}_{j1}^\dagger \right)^{-1} \mathbf{U}_1 \tag{46}$$

With $q_{jk}$ denoting the $jk$-th entry of $\mathbf{Q}$, we have:

$$q_{jj} = \mathbf{u}_{1j}^\dagger \left( N\bar{\sigma}^2 \mathbf{I} + \sum_{j=2}^{n+1} \gamma_{1j} \tilde{\mathbf{H}}_{j1} \mathbf{P}_j \tilde{\mathbf{H}}_{j1}^\dagger \right)^{-1} \mathbf{u}_{1j} \tag{47}$$

where $\mathbf{u}_{1k}$ is the $k$-th column of $\mathbf{U}_1$.

We can write (45) as:

$$C_1 = \log_2 \left| \mathbf{I} + \gamma_{11} \mathbf{\Sigma}_1 \mathbf{P}_1 \mathbf{\Sigma}_1^\dagger \mathbf{Q} \right|. \tag{48}$$

$\mathbf{\Sigma}_1$ contains the singular values of the channel matrix of the representative link with $\lambda_{1j}$ equal to the square of the $j$-th largest singular value. Thus, from (42) the $j$-th diagonal entry of $\mathbf{I} + \gamma_{11} \mathbf{\Sigma}_1 \mathbf{P}_1 \mathbf{\Sigma}_1^\dagger \mathbf{Q}$ is $1 + \gamma_{11} \lambda_{1j} P_{1j} q_{jj}$. Hence, by the Hadamard inequality (see e.g. [28]) we can bound (48) as follows:

$$C_1 \leq \log_2 \left( \prod_{j=1}^{K} \left( 1 + \gamma_{11} \lambda_{1j} P_{1j} q_{jj} \right) \right) = \sum_{j=1}^{M} \log_2 \left( 1 + \gamma_{11} \lambda_{1j} P_{1j} q_{jj} \right) \tag{49}$$

where the last step uses the fact that $P_{1j} = 0$ for $j > M$. Substituting (47) yields:

$$C_1 \leq \sum_{j=1}^{M} \log_2 \left( 1 + \gamma_{11} \lambda_{1j} P_{1j} \mathbf{u}_{1j}^\dagger \left( N\bar{\sigma}^2 \mathbf{I} + \sum_{j=2}^{n+1} \gamma_{1j} \tilde{\mathbf{H}}_{j1} \mathbf{P}_j \tilde{\mathbf{H}}_{j1}^\dagger \right)^{-1} \mathbf{u}_{1j} \right) \tag{50}$$

Using some matrix manipulations, (50) can be written as:

$$C_1 \leq \sum_{j=1}^{M} \log_2 \left( 1 + \gamma_{11} \lambda_{1j} P_{1j} \mathbf{u}_{1j}^\dagger \left( N\bar{\sigma}^2 \mathbf{I} + \mathbf{K}_1 \mathbf{\Phi}_1 \mathbf{K}_1^\dagger \right)^{-1} \mathbf{u}_{1j} \right) \tag{51}$$

where $\mathbf{\Phi}_1$ is given in (7) and

$$\mathbf{K}_1 = \left[ \tilde{\mathbf{H}}_{21}' \ \tilde{\mathbf{H}}_{31}' \cdots \ \tilde{\mathbf{H}}_{(n+1)1}' \right] \tag{52}$$

where the $N \times M$ matrix $\tilde{\mathbf{H}}'_{i1}$ comprises the first $M$ columns of $\tilde{\mathbf{H}}_{i1}$.

Note that $\boldsymbol{\Phi}_1$ is a diagonal matrix containing the received powers from each stream transmitted by each interferer and $\mathbf{K}_1$ is a matrix of IID $\mathcal{CN}(0,1)$ entries.

*2) Lower Bound on Spectral Efficiency:* Since the transmit covariance matrix of the $j$-th transmitter is $\mathbf{V}_j \mathbf{P}_j \mathbf{V}_j^\dagger$, the vector of received samples at the antennas of the representative receiver at a given sampling time can be written as follows:

$$\mathbf{y} = \mathbf{H}_{11} \mathbf{V}_1 \mathbf{P}_1^{\frac{1}{2}} \mathbf{x}_1 + \sum_{j=2}^{n+1} \sqrt{\gamma_{1j}} \tilde{\mathbf{H}}_{j1} \mathbf{P}_j^{\frac{1}{2}} \mathbf{x}_j + \mathbf{n} \tag{53}$$

where $\mathbf{x}_j$ contains unit variance samples from $\mathcal{T}_j$ in its first $M$ entries with the remainder being zeros, and $\mathbf{n}$ is an $N \times 1$ vector of IID $\mathcal{CN}(0, \bar{\sigma}^2 N)$ entries representing the thermal noise.

We shall find the lower bound by using a specific, suboptimal procedure where the representative receiver decodes each stream from the representative transmitter individually. When decoding the $i$-th stream from the representative transmitter, the receiver will use a fraction of its degrees of freedom to completely null the interference from the other data streams of the representative transmitter.

Suppose that the receiver multiples $\mathbf{y}$ by $\mathbf{U}_1^\dagger$, which yields:

$$\begin{aligned}
\bar{\mathbf{y}} = \mathbf{U}_1^\dagger \mathbf{y} &= \mathbf{U}_1^\dagger \mathbf{H}_{11} \mathbf{V}_1 \mathbf{P}_1^{\frac{1}{2}} \mathbf{x}_1 + \mathbf{U}_1^\dagger \sum_{j=2}^{n+1} \sqrt{\gamma_{1j}} \tilde{\mathbf{H}}_{j1} \mathbf{P}_j^{\frac{1}{2}} \mathbf{x}_j + \mathbf{U}_1^\dagger \mathbf{n} \\
&= \mathbf{U}_1^\dagger \mathbf{U}_1 \boldsymbol{\Sigma}_1 \mathbf{V}_1^\dagger \mathbf{V}_1 \mathbf{P}_1^{\frac{1}{2}} \mathbf{x}_1 + \sum_{j=2}^{n+1} \sqrt{\gamma_{1j}} \mathbf{U}_1^\dagger \tilde{\mathbf{H}}_{j1} \mathbf{P}_j^{\frac{1}{2}} \mathbf{x}_j + \mathbf{U}_1^\dagger \mathbf{n} \\
&= \boldsymbol{\Sigma}_1 \mathbf{P}_1^{\frac{1}{2}} \mathbf{x}_1 + \sum_{j=2}^{n+1} \sqrt{\gamma_{1j}} \bar{\mathbf{H}}_{j1} \mathbf{P}_j^{\frac{1}{2}} \mathbf{x}_j + \bar{\mathbf{n}} \\
&= \bar{\mathbf{x}}_1 + \sum_{j=2}^{n+1} \sqrt{\gamma_{1j}} \bar{\mathbf{H}}_{j1} \mathbf{P}_j^{\frac{1}{2}} \mathbf{x}_j + \bar{\mathbf{n}}
\end{aligned} \tag{54}$$

where $\bar{\mathbf{H}}_{j1} = \mathbf{U}_1^\dagger \tilde{\mathbf{H}}_{j1}$ is a matrix with IID $\mathcal{CN}(0,1)$ and $\bar{\mathbf{n}} = \mathbf{U}_1^\dagger \mathbf{n}$ contains IID $\mathcal{CN}(0, \bar{\sigma}^2 N)$ entries. Note that

$$\bar{\mathbf{x}}_1 = \boldsymbol{\Sigma}_1 \mathbf{P}_1^{\frac{1}{2}} \mathbf{x}_1 = \begin{pmatrix} \sqrt{\lambda_{11} P_{11}} x_{11} \\ \vdots \\ \sqrt{\lambda_{1M} P_{1M}} x_{1M} \\ 0 \\ \vdots \\ 0 \end{pmatrix} \tag{55}$$

where the zero entries appear because $P_{1j} = 0$ for $j > M$ and $x_{1j}$ is the $j$th data sample of the representative transmitter, i.e., the $j$th entry of the vector $\mathbf{x}_1$. Note from (54) and (55) that the data samples from the representative transmitter are parallelized in $\bar{\mathbf{y}}$ so each sample of $\bar{\mathbf{y}}$ contains information from a single stream of the representative transmitter.

To decode the $i$th data sample from the representative transmitter the receiver constructs a vector $\tilde{\mathbf{y}}_i$ by discarding samples that contain information from the remaining $M - 1$ streams of the representative transmitter. The $(N - M + 1) \times 1$ vector $\tilde{\mathbf{y}}_i$ is defined as follows:

$$\tilde{\mathbf{y}}_i = \begin{pmatrix} \bar{y}_i \\ \bar{y}_{(M+1)} \\ \vdots \\ \bar{y}_N \end{pmatrix} \tag{56}$$

where $\bar{y}_j$ is the $j$th entry of the vector $\bar{\mathbf{y}}$. Thus, $\check{\mathbf{y}}_i$ can be written as follows:

$$\check{\mathbf{y}}_i = \begin{pmatrix} \sqrt{\lambda_{1i}P_{1i}}x_i \\ 0 \\ \vdots \\ 0 \end{pmatrix} + \sum_{j=2}^{n+1} \sqrt{\gamma_{1j}}\check{\mathbf{H}}_{ji}\mathbf{P}_j^{\frac{1}{2}}\mathbf{x}_j + \check{\mathbf{n}} \tag{57}$$

where $\check{\mathbf{H}}_{ji}$ is an $(N - M + 1) \times K$ matrix of IID $\mathcal{CN}(0,1)$, which equals the matrix $\bar{\mathbf{H}}_{j1}$ with the rows $1, \cdots, i-1, i+1, \cdots, M$ removed. The $(N - M + 1) \times 1$ vector $\check{\mathbf{n}}$ contains IID $\mathcal{CN}(0, \bar{\sigma}^2 N)$ noise samples.

Suppose now that the receiver picks an $(N - M + 1) \times (N - M + 1)$ matrix $\hat{\mathbf{U}}$ with uniform probability from the group of all unitary matrices and multiples it with $\check{\mathbf{y}}_i$. This yields the following:

$$\hat{\mathbf{y}}_i = \mathbf{U}\check{\mathbf{y}}_i = \sqrt{\lambda_{1i}P_{1i}}\hat{\mathbf{u}}_i x_i + \sum_{j=2}^{n+1} \sqrt{\gamma_{1j}}\hat{\mathbf{H}}_{ji}\mathbf{P}_j^{\frac{1}{2}}\mathbf{x}_j + \hat{\mathbf{n}}. \tag{58}$$

where $\hat{\mathbf{u}}_i$ is the $i$th column of $\hat{\mathbf{U}}$. Since $\hat{\mathbf{U}}$ is unitary, $\hat{\mathbf{H}}_{j1} = \hat{\mathbf{U}}\check{\mathbf{H}}_{j1}$ is still an $(N - M + 1) \times K$ matrix of IID $\mathcal{CN}(0,1)$ entries and $\hat{\mathbf{n}} = \hat{\mathbf{U}}\check{\mathbf{n}}$ is an $(N - M + 1) \times 1$ vector of IID $\mathcal{CN}(0, \bar{\sigma}^2 N)$ noise samples.

The receiver then uses a linear MMSE receiver on the vector $\hat{\mathbf{y}}_i$ to detect $x_i$ for which the SINR is well known (e.g. see [7]):

$$\text{SINR}_i = \lambda_{1i}P_{1i}\hat{\mathbf{u}}_i^\dagger \left( \sum_{j=2}^{n+1} \gamma_{1j}\hat{\mathbf{H}}_{ji}\mathbf{P}_j\hat{\mathbf{H}}_{ji}^\dagger + N\bar{\sigma}^2\mathbf{I} \right)^{-1} \hat{\mathbf{u}}_i = \lambda_{1i}P_{1i}\hat{\mathbf{u}}_i^\dagger \left( \hat{\mathbf{K}}_i\mathbf{\Phi}_1\hat{\mathbf{K}}_i^\dagger + N\bar{\sigma}^2\mathbf{I} \right)^{-1} \hat{\mathbf{u}}_i \tag{59}$$

where

$$\hat{\mathbf{K}}_i = \begin{bmatrix} \hat{\mathbf{H}}'_{2i} & \hat{\mathbf{H}}'_{3i} \cdots & \hat{\mathbf{H}}'_{(n+1)i} \end{bmatrix}. \tag{60}$$

where $\hat{\mathbf{H}}'_{ji}$ is the matrix comprising the first $M$ columns of $\hat{\mathbf{H}}_{ji}$ and $\mathbf{\Phi}_1$ is defined in (7).

Since all transmissions use Gaussian codebooks, each stream can support a spectral efficiency of

$$\log_2 \left( 1 + \lambda_{1i}P_{1i}\hat{\mathbf{u}}_i^\dagger \left( \hat{\mathbf{K}}_i\mathbf{\Phi}_1\hat{\mathbf{K}}_i^\dagger + N\bar{\sigma}^2\mathbf{I} \right)^{-1} \hat{\mathbf{u}} \right) \tag{61}$$

Summing the contributions of $M$ streams gives the lower bound.

### B. Proof of Lemma 2 on the Limiting SINR per stream.

Note that $\mathbf{w}_j$ in (16) is an isotropic random vector with unit norm and (e.g. see [22], [29], [30] or [31] Appendix A.1.) can be expressed as [3] :

$$\mathbf{w}_j = \frac{1}{||\mathbf{g}_j||}\mathbf{g}_j \tag{62}$$

where the entries of $\mathbf{g}_j$ are distributed as IID $\mathcal{CN}(0,1)$, which means that (16) can be written as:

$$\text{SINR}_j = \frac{1}{\frac{1}{N}||\mathbf{g}_j||^2}\gamma_{1j}P_{1j}\frac{1}{N}\lambda_{1j}\frac{1}{N}\mathbf{g}_j^\dagger \left( \bar{\sigma}^2\mathbf{I} + \frac{1}{N}\mathbf{K}_1\mathbf{\Phi}_1\mathbf{K}_1^\dagger \right)^{-1} \mathbf{g}_j \tag{63}$$

If the e.d.f of the entries of $\mathbf{\Phi}_1$ converges with probability 1 to a limiting function $H(\tau)$, the factor

$$\frac{1}{N}\mathbf{g}_j^\dagger \left( \bar{\sigma}^2\mathbf{I} + \frac{1}{N}\mathbf{K}_1\mathbf{\Phi}_1\mathbf{K}_1^\dagger \right)^{-1} \mathbf{g}_j \tag{64}$$

was shown by [23] to converge with probability 1 to an asymptotic limit as $n \to \infty$. The following lemma proved in Appendix C shows that the e.d.f. of the entries of $\mathbf{\Phi}_1$ indeed converges with probability 1 to an asymptotic limit.

---

[3]Note that expressing isotropic vectors in this manner was used in [22] in their analysis of the fluctuation about the mean of the SINR of random CDMA systems with signature vectors comprising IID entries.

*Lemma 4:* Let $H_n(x)$ denote the e.d.f of the interference powers for any given $n$, i.e. $H_n(x)$ is the e.d.f of the entries of the diagonal matrix $\boldsymbol{\Phi}_1$. Then as $n \to \infty$ in the manner of Theorem 1,

$$H_n(x) \to H(x) = \frac{1}{M} \sum_{j=1}^{M} \int f_i(\tau) \Psi(x/\tau) \, d\tau \tag{65}$$

From Lemma 4 and the main result of [23], the term $\frac{1}{N} \mathbf{g}_j^\dagger \left( \bar{\sigma}^2 \mathbf{I} + \mathbf{K}_1 \boldsymbol{\Phi}_1 \mathbf{K}_1^\dagger \right)^{-1} \mathbf{g}_j$ converges with probability 1 to an asymptotic limit which we define as $\beta$. From [21] $\frac{1}{N} \mathbf{g}^\dagger \left( \bar{\sigma}^2 \mathbf{I} + \mathbf{K}_1 \boldsymbol{\Phi}_1 \mathbf{K}_1^\dagger \right)^{-1} \mathbf{g}$ converges *in probability* to $\beta$ which is a unique solution for $\beta(z)$ in the equation:

$$z\beta(z) + 1 = \beta(z) c \int_0^\infty \frac{\tau \, dH(\tau)}{1 + \tau \beta(z)} \tag{66}$$

when $z = -\bar{\sigma}^2$. Since [23] proves convergence with probability 1 to an asymptotic limit as $n \to \infty$, and [21] proves convergence in probability to the solution to (66), we conclude that $\frac{1}{N} \mathbf{g}_j^\dagger \left( \bar{\sigma}^2 \mathbf{I} + \mathbf{K}_1 \boldsymbol{\Phi}_1 \mathbf{K}_1^\dagger \right)^{-1} \mathbf{g}_j$ converges with probability 1 to the solution to (66).

Additionally, note that

$$\frac{1}{\frac{1}{N} ||\mathbf{g}||^2} \to 1$$

with probability 1 by the strong law of large numbers.

If $K \to \infty$ as $N \to \infty$ with $K/N = d$, then $\frac{1}{N} \lambda_{1j}$ for $j = 1, 2, \cdots, M$ is known to converge with probability 1 to an asymptotic limit $\lambda_j^* = (1 + \sqrt{d})^2$ for $j = 1, 2, \cdots, M$ (e.g. see [32]). If $K$ is a finite constant, by the strong law of large numbers $\frac{1}{N} \lambda_{1j}$ for $j = 1, 2, \cdots, M$ converges with probability 1 to unity.

Hence, each random term in (16) converges with probability 1 to a non-random value. It is known that for sequences of random variables $X_n$ and $Y_n$, if $X_n \to X$ and $Y_n \to Y$ with probability 1, then $X_n Y_n \to XY$ with probability 1 as well (e.g. see Theorem 5.21 in [27]). Hence, (16) converges with probability 1 to

$$\gamma_1 P_{1j} \lambda_j^* \beta \tag{67}$$

### C. Proof of Lemma 4 on the convergence of the Interference Powers

Let $p_{ij} = \gamma_i P_{ij}$, i.e. $p_{ij}$ is the product of path-loss to the representative receiver and the transmit power on the $j$-th stream of the $i$-th transmitting node. Recall that $N, n \to \infty$ with $n/N = c/M > 0$. Thus we have the limiting e.d.f. of the received interference powers:

$$H(x) = \lim_{n \to \infty} H_n(x) = \lim_{n \to \infty} \frac{1}{M} \sum_{j=1}^{M} \frac{1}{n} \sum_{i=2}^{n+1} 1_{\{p_{ij} \le x\}} \tag{68}$$

$$= \frac{1}{M} \sum_{j=1}^{M} E\left[ 1_{\{p_{ij} \le x\}} \right] \quad \text{with probability 1} \tag{69}$$

$$= \frac{1}{M} \sum_{j=1}^{M} Pr\{p_{ij} < x\} \tag{70}$$

where the step from (68) to (69) follows from the strong-law-of-large-numbers.

Since $p_{ij}$ are distributed identically for all $i$,

$$Pr\{p_{ij} < x\} = \int f_j(\tau) Pr\{p_{ij} < x | P_{ij} = \tau\} d\tau \tag{71}$$

Substituting $p_{ij} = \gamma_i P_{ij}$ yields:

$$Pr\{p_{ij} < x\} = \int f_j(\tau) Pr\{\gamma_i \tau < x | P_{ij} = \tau\} d\tau \tag{72}$$

$$= \int f_j(\tau) \Psi(x/\tau) d\tau \tag{73}$$

Thus we have:

$$H(x) = \lim_{n \to \infty} H_n(x) = \int f_j(\tau) \Psi(x/\tau) d\tau \tag{74}$$

### D. Proof of Theorem 2 on the limiting SINR for the spatially distributed network model

Recall that the path losses from each interferer and the thermal noise are scaled by $N^{\alpha/2}$ for this model. The SINR on the $j$-th stream from the representative transmitter can be bounded as follows:

$$\gamma_1 P_{1j} \lambda_{1j} \hat{\mathbf{u}}_j^\dagger \left( N\bar{\sigma}^2 \mathbf{I} + \hat{\mathbf{K}}_j \mathbf{\Phi}_1 \hat{\mathbf{K}}_j^\dagger \right)^{-1} \hat{\mathbf{u}}_j \le \eta_{Nj} \le \gamma_1 P_{1j} \lambda_{1j} \mathbf{u}_{1j}^\dagger \left( N\bar{\sigma}^2 \mathbf{I} + \mathbf{K}_1 \mathbf{\Phi}_1 \mathbf{K}_1^\dagger \right)^{-1} \mathbf{u}_{1j} \tag{75}$$

where $\hat{\mathbf{u}}_j$, $\mathbf{u}_j$, $\hat{\mathbf{K}}_j$ and $\mathbf{K}_1$ are as defined in Lemma 1. Recall that the upper bound is attained if the $M$ streams from the representative transmitter do not interfere with each other and the lower bound is attained by perfectly nulling the interference of streams $1, \cdots, (i-1), (i+1), \cdots, M$ from the representative transmitter.

If this new system with the scaled path-losses and noise power meets the conditions of Lemma 2, then the upper and lower bounds in (75) will converge to the same asymptotic limit, implying that the normalized SINR $\eta_{Nj}$ converges to that limit as well. The rest of this section is devoted to showing that the system model meets the requirements of Lemma 2 and to finding the limiting value of the upper and lower bounds in (75).

We start by showing that the empirical distribution of received interference powers converges with probability 1 to an asymptotic limit $H(\tau)$. For this network model and a given $n, N$, and $R$, let the CDF of the path-losses be denoted by $\Psi_N(x)$. Note that the interference power is scaled by $N^{\alpha/2}$ here.

$$\begin{aligned}
\Psi_N(x) &= \Pr\{N^{\alpha/2} G_t r_i^{-\alpha} < x\} \\
&= \Pr\left\{ r_i > \left( \frac{x N^{-\alpha/2}}{G_t} \right)^{-\frac{1}{\alpha}} \right\} \\
&= \frac{R^2 - \left( \frac{x N^{-\alpha/2}}{G_t} \right)^{-\frac{2}{\alpha}}}{R^2} \mathbf{1}_{\left\{ 0 < \left( \frac{x N^{-\alpha/2}}{G_t} \right)^{-\frac{1}{\alpha}} < R \right\}}
\end{aligned} \tag{76}$$

Substituting (1), $c = nM/N$, and $b = \left( \frac{N\pi\rho}{n} \right)^{\alpha/2}$

$$\begin{aligned}
\Psi_N(x) &= \frac{\frac{n}{\pi\rho} - \left( \frac{x N^{-\alpha/2}}{G_t} \right)^{-\frac{2}{\alpha}}}{\frac{n}{\pi\rho}} \mathbf{1}_{\left\{ 0 < \left( \frac{x N^{-\alpha/2}}{G_t} \right)^{-\frac{1}{\alpha}} < \sqrt{\frac{n}{\pi\rho}} \right\}} \\
&= 1 - \frac{N\pi\rho}{n} \left( \frac{x}{G_t} \right)^{-\frac{2}{\alpha}} \mathbf{1}_{\left\{ 0 < \left( \frac{x}{G_t} \right)^{-\frac{1}{\alpha}} < \sqrt{\frac{n}{N\pi\rho}} \right\}} \\
&= 1 - \frac{\pi\rho M}{c} \left( \frac{x}{G_t} \right)^{-\frac{2}{\alpha}} \mathbf{1}_{\{G_t b < x < \infty\}} = \Psi(x)
\end{aligned}$$

Hence $\Psi_N(x)$ is independent of $N$ (although it depends on $n/N$ which is a constant) and Lemma 2 holds for the spatially distributed model with scaled interference powers. The remaining steps are to find $H(x)$ and its derivative, which are used to evaluate (16), which gives the limiting normalized SINR.

$$H(x) = \frac{1}{M} \sum_{j=1}^{M} \int f_i(\tau) \Psi(x/\tau) \, d\tau$$

$$= \frac{1}{M} \sum_{j=1}^{M} \int f_i(\tau) \left( 1 - \frac{\pi \rho M}{c} \left( \frac{x/\tau}{G_t} \right)^{-\frac{2}{\alpha}} 1_{\{G_t b < x/\tau < \infty\}} \right) d\tau$$

$$= 1 - \sum_{j=1}^{M} \int_0^{\frac{x}{b G_t}} f_i(\tau) \left( \frac{\pi \rho}{c} \left( \frac{x/\tau}{G_t} \right)^{-\frac{2}{\alpha}} \right) d\tau$$

$$= 1 - \frac{\pi \rho G_t^{\frac{2}{\alpha}}}{c x^{\frac{2}{\alpha}}} \sum_{j=1}^{M} \int_0^{\frac{x}{b G_t}} f_i(\tau) \tau^{\frac{2}{\alpha}} \, d\tau$$

$$= 1 - \frac{\pi \rho G_t^{\frac{2}{\alpha}}}{c x^{\frac{2}{\alpha}}} \sum_{j=1}^{M} \int_0^{\infty} f_i(\tau) \tau^{\frac{2}{\alpha}} \, d\tau + \frac{\pi \rho G_t^{\frac{2}{\alpha}}}{c x^{\frac{2}{\alpha}}} \sum_{j=1}^{M} \int_{\frac{x}{b G_t}}^{\infty} f_i(\tau) \tau^{\frac{2}{\alpha}} \, d\tau$$

$$= 1 - \frac{\pi \rho G_t^{\frac{2}{\alpha}}}{c x^{\frac{2}{\alpha}}} \sum_{j=1}^{M} E[P_j^{\frac{2}{\alpha}}] + \frac{\pi \rho G_t^{\frac{2}{\alpha}}}{c x^{\frac{2}{\alpha}}} \sum_{j=1}^{M} \int_{\frac{x}{b G_t}}^{\infty} f_i(\tau) \tau^{\frac{2}{\alpha}} \, d\tau$$

The derivative of $H(x)$ is:

$$\frac{dH(x)}{dx} = \frac{2\pi \rho G_t^{\frac{2}{\alpha}}}{\alpha c x^{1+\frac{2}{\alpha}}} \sum_{j=1}^{M} E[P_j^{\frac{2}{\alpha}}] - \frac{2\pi \rho G_t^{\frac{2}{\alpha}}}{\alpha c x^{1+\frac{2}{\alpha}}} \sum_{j=1}^{M} \int_{\frac{x}{b G_t}}^{\infty} f_i(\tau) \tau^{\frac{2}{\alpha}} \, d\tau \tag{77}$$

Substituting (77) into (9) and scaling the thermal noise power by $N^{\alpha/2}$ such that $\bar{\sigma}^2 = \sigma^2$ yields:

$$-\sigma^2 \beta + 1 = \beta c \int_0^{\infty} \frac{x}{1 + x\beta} \left[ \frac{2\pi \rho G_t^{\frac{2}{\alpha}}}{\alpha c x^{1+\frac{2}{\alpha}}} \sum_{j=1}^{M} E[P_j^{\frac{2}{\alpha}}] - \frac{2\pi \rho G_t^{\frac{2}{\alpha}}}{\alpha c x^{1+\frac{2}{\alpha}}} \sum_{j=1}^{M} \int_{\frac{x}{b G_t}}^{\infty} f_i(\tau) \tau^{\frac{2}{\alpha}} \right] dx$$

$$= \frac{2\pi \rho \beta c G_t^{\frac{2}{\alpha}}}{\alpha c} \sum_{j=1}^{M} E[P_j^{\frac{2}{\alpha}}] \int_0^{\infty} \frac{x^{-\frac{2}{\alpha}}}{1 + x\beta} \, dx - \frac{2\pi \rho G_t^{\frac{2}{\alpha}}}{\alpha c} \sum_{j=1}^{M} \int_0^{\infty} \frac{x^{-\frac{2}{\alpha}}}{1 + x\beta} \int_{\frac{x}{b G_t}}^{\infty} f_i(\tau) \tau^{\frac{2}{\alpha}} \, d\tau \, dx \tag{78}$$

From the proof of Lemma 1 in [11][4],

$$\int_0^{\infty} \frac{x^{-\frac{2}{\alpha}}}{1 + x\beta} \, dx = \beta^{\frac{2}{\alpha}-1} \pi csc\left( \frac{2\pi}{\alpha} \right) \tag{79}$$

Substituting (79) into (78):

$$-\sigma^2 \beta + 1 = \frac{2\pi \rho \beta^{\frac{2}{\alpha}} c G_t^{\frac{2}{\alpha}}}{\alpha c} \sum_{j=1}^{M} E[P_j^{\frac{2}{\alpha}}] \pi csc\left( \frac{2\pi}{\alpha} \right) - \frac{2\pi \rho G_t^{\frac{2}{\alpha}}}{\alpha c} \sum_{j=1}^{M} \int_0^{\infty} \frac{x^{-\frac{2}{\alpha}}}{1 + x\beta} \int_{\frac{x}{b G_t}}^{\infty} f_i(\tau) \tau^{\frac{2}{\alpha}} \, d\tau \, dx.$$

Rearranging terms yields (29) which completes the proof.

---

[4]Note that the lower limit of the integral in Lemma 1 of [11] is greater than zero, however its proof clearly allows the lower limit to be zero

*E. Approximating the Mean Spectral Efficiency by the Asymptotic Spectral Efficiency*

Consider the deviation of the mean spectral efficiency from the asymptotic spectral efficiency on the i-th stream. If we assume that $E\left[\log_2\left(\eta_{Ni}\right)\right]$ is bounded for all $N$, then by the bounded-convergence theorem (see e.g. [27]),

$$\left|E\left[\log_2\left(1 + N^{\frac{\alpha}{2}}\eta_{Ni}\right)\right] - \log_2\left(1 + N^{\frac{\alpha}{2}}\eta_i\right)\right| = \left|E\left[\log_2\left(\frac{N^{-\frac{\alpha}{2}} + \eta_{Ni}}{N^{-\frac{\alpha}{2}} + \eta_i}\right)\right]\right| \rightarrow 0. \tag{80}$$

This implies that the deviation of the mean spectral efficiency from its asymptotic value decays to zero, i.e.:

$$\sum_{i=1}^{M} E\left[\log_2\left(1 + N^{\frac{\alpha}{2}}\eta_{Ni}\right)\right] - \sum_{i=1}^{M} \log_2\left(1 + N^{\frac{\alpha}{2}}\eta_i\right) \rightarrow 0. \tag{81}$$

Thus for large $N$, the mean spectral efficiency is well approximated by the asymptotic spectral efficiency.

To show that $E\left[\log_2\left(\eta_{Ni}\right)\right]$ is bounded for all $N$, consider:

$$\left|E\left[\log_2\left(\eta_{Ni}\right)\right]\right| \leq E\left[\left|\log_2\left(\eta_{Ni}\right)\right|\right] = \int_0^\infty \Pr\left\{\left|\log_2\left(\eta_{Ni}\right)\right| > x\right\}\, dx$$

$$= \int_0^\infty \Pr\left\{\eta_{Ni} \geq 2^x\right\}\, dx + \int_0^\infty \Pr\left\{\eta_{Ni} < 2^{-x}\right\}\, dx$$

The first term on the RHS of (82) is bounded because $\eta_{Ni}$ is bounded from above. Now, consider the second term on the RHS:

$$\Pr\left\{\eta_{Ni} < 2^{-x}\right\} \leq \Pr\left\{\bar{\eta}_{Ni} < 2^{-x}\right\} \tag{82}$$

where $\bar{\eta}_{Ni}$ is the normalized SINR of the representative link in an infinite wireless network with $N$ antennas at the representative receiver, no CSI at the transmitters, and any suboptimal linear receiver.

Consider a suboptimal receiver which uses only one of its antennas selected at random if $N \leq \frac{1}{\theta}\lceil\frac{\alpha}{2}\rceil$, for some $0 < \theta < \frac{1}{2}$. In this case from [33],

$$\Pr\left\{\bar{\eta}_{Ni} < 2^{-x}\right\} = 1 - \exp\left(-G_1 N 2^{-x} - G_2 N 2^{-\frac{2x}{\alpha}}\right)$$

$$\leq 1 - \exp\left(-\left(NG_1 + NG_2\right)2^{-\frac{2x}{\alpha}}\right) \tag{83}$$

where $G_1$ and $G_2$ are positive parameters independent of $x$. The integral of (83) w.r.t. $x$ is finite.

If $N > \frac{1}{\theta}\lceil\frac{\alpha}{2}\rceil$, the receiver uses the partial-zero-forcing algorithm of [16] with $k = \theta N$ degrees of freedom used for zero-forcing. In this case, we have from [16]:

$$\Pr\left\{\bar{\eta}_{Ni} < 2^{-x}\right\} < \frac{N^{\frac{\alpha}{2}}2^{-x}\left(G_3\left(\theta N - \lceil\frac{\alpha}{2}\rceil\right)^{1-\frac{\alpha}{2}} + G_4 N^{1-\alpha/2}\sigma^2\right)}{(1-\theta)N - 1}$$

$$= \frac{N^{\frac{\alpha}{2}-1}2^{-x}G_3\left(\theta N - \lceil\frac{\alpha}{2}\rceil\right)^{1-\frac{\alpha}{2}} + G_4\sigma^2}{(1-\theta) - 1/N}$$

$$\leq \frac{N^{\frac{\alpha}{2}-1}2^{-x}G_3\left(\theta N - \lceil\frac{\alpha}{2}\rceil\right)^{1-\frac{\alpha}{2}} + G_4\sigma^2}{1 - \left(1 + \frac{1}{\lceil\frac{\alpha}{2}\rceil}\right)\theta}$$

$$= \frac{2^{-x}G_3\left(\theta - \frac{1}{N}\lceil\frac{\alpha}{2}\rceil\right)^{1-\frac{\alpha}{2}} + G_4\sigma^2}{1 - \left(1 + \frac{1}{\lceil\frac{\alpha}{2}\rceil}\right)\theta}$$

$$\leq \frac{2^{-x}G_3\left(\theta - \frac{1}{\lceil\frac{1}{\theta}\lceil\frac{\alpha}{2}\rceil\rceil}\lceil\frac{\alpha}{2}\rceil\right)^{1-\frac{\alpha}{2}} + G_4\sigma^2}{1 - \left(1 + \frac{1}{\lceil\frac{\alpha}{2}\rceil}\right)\theta} \tag{84}$$

where $G_3, G_4$ are positive terms independent of $N$ and $x$.

The integral of the RHS of (84) is clearly finite which implies that $E\left[\left|\log_2\left(\eta_{Ni}\right)\right|\right]$ is finite confirming (81).

*F. Asymptotic Empirical Distribution Function of the Eigenvalues of Wishart Matrices*

It is known that (e.g. see [32]):

$$\frac{dF_d(y)}{dy} = \max(0, 1-d)\delta(y) - \frac{1}{2\pi y}\sqrt{(a_2-y)(y-a_1)}1_{\{a_1 < y < a_2\}} \tag{85}$$

where $a_1 = (1-\sqrt{d})^2$ and $a_2 = (1+\sqrt{d})^2$. Taking the integral with respect to $y$ from 0 to $x$, we find that for $a_1 < y < a_2$

$$F_d(y) = \frac{1}{8}(a_1+a_2) - \frac{1}{4}\sqrt{a_1 a_2} + \frac{1}{2\pi}\sqrt{(a_2-y)(y-a_1)} + \frac{1}{4\pi}(a_1+a_2)\arcsin\left(\frac{a_1+a_2-2y}{a_1-a_2}\right)$$
$$+ \frac{1}{2\pi}\sqrt{a_1 a_2}\arctan\left(\frac{2\,a_1 a_2 - a_2 y - a_1 y}{2\sqrt{a_1 a_2 (a_2-y)(y-a_1)}}\right) \tag{86}$$

and

$$F_d(x) = \begin{cases} \max(0, 1-d) & \text{if } 0 \le x \le a_1 \\ 1 & \text{if } a_2 \le x \end{cases} \tag{87}$$

*G. Asymptotic Spectral Efficiency for Constant Path-loss and the Two-Class Model*

For simplicity, we express (25) as

$$T_1 \beta^3 + T_2 \beta^2 + T_3 \beta - 1 = 0 \tag{88}$$

where

$$T_1 = \bar{\sigma}^2 P_1 P_2 \gamma^2$$
$$T_2 = \frac{(1-q)cP_1 P_2 \gamma^2}{M} + cq P_1 P_2 \gamma^2 + (\bar{\sigma}^2 - P_1 \gamma)P_2 \gamma + \bar{\sigma}^2 P_1 \gamma$$
$$T_3 = -P_2 \gamma - P_1 \gamma + \bar{\sigma}^2 + \frac{(1-q)cP_2 \gamma}{M} + qc P_1 \gamma$$

Additionally, we define the following to simplify notation:

$$T_4 = 2T_2^3 - 9T_1 T_2 T_3 - 27T_1^2$$
$$T_5 = \sqrt{T_4^2 - 4(T_2^2 - 3T_1 T_3)^3}$$

The solution to (25) is found using the standard formula for the roots of a cubic polynomial, which yields:

$$\beta_{2c} = -\frac{T_2}{3T_1} + \frac{1-j\sqrt{3}}{6T_1}\left(\frac{T_4+T_5}{2}\right)^{\frac{1}{3}} + \frac{1+j\sqrt{3}}{6T_1}\left(\frac{T_4-T_5}{2}\right)^{\frac{1}{3}} \tag{89}$$

*H. Proof of Lemma 3 Used to Simplify Limiting Mean Spectral Efficiency*

Note that $\beta$ must monotonically decrease with $b = \left(\frac{\pi \rho N}{n}\right)^{\alpha/2}$, and that the maximum transmit power per stream is $P_M$ since the total transmit power is bounded by $P_M$. Thus, $f_i(x) = 0$ for $x > P_M$ and so for any $i$,

$$\int_0^\infty \frac{\tau^{-\frac{2}{\alpha}}}{1+\tau\beta}\int_{\frac{\tau}{G_t b}}^\infty f_i(x)x^{\frac{2}{\alpha}}dx\,d\tau$$

$$= \int_0^{bP_M} \frac{\tau^{-\frac{2}{\alpha}}}{1+\tau\beta}\int_{\frac{\tau}{G_t b}}^\infty f_i(x)x^{\frac{2}{\alpha}}dx\,d\tau \tag{90}$$

$$\le \int_0^{bP_M} \frac{\tau^{-\frac{2}{\alpha}}}{1+\tau\beta}\int_{\frac{\tau}{G_t b}}^\infty f_i(x)P_M^{\frac{2}{\alpha}}dx\,d\tau \tag{91}$$

$$\le \int_0^{bP_M} \frac{\tau^{-\frac{2}{\alpha}}}{1+\tau\beta}P_M^{\frac{2}{\alpha}}\,d\tau \tag{92}$$

which goes to zero as $b \to 0$. The step from (90) to (91) is because the maximum transmit power is $P_M$ and (92) follows because PDFs integrate to unity.

## References

[1] S. Govindasamy, D. W. Bliss, and D. Staelin, "Spectral-Efficiency of Multi-antenna Links in Ad-hoc Wireless Networks with Limited Tx CSI," *Conference Record for the 43rd Asilomar Conf. on Signals, Systems and Computers*, 2009.

[2] D. Tse and P. Viswanath, *Fundamentals of Wireless Communication*. Cambridge University Press, 2005.

[3] J. Winters, "Optimum combining in digital mobile radio with cochannel interferernce," *IEEE Journal on Selected Areas of Communications*, vol. SAC-2, July 1984.

[4] G. J. Foschini, "Layered space-time architecture for wireless communication in a fading environment when using multi- element antennas," *AT&T Bell Labs Tech. Journal*, vol. 2, 1996.

[5] E. Telatar, "Capacity of multi-antenna Gaussian channels," *European Transactions on Telecommuncations ETT*, vol. 10, no. 6, pp. 585–595, 1999.

[6] L. Zheng and D. Tse, "Diversity and multiplexing:a fundamental tradeoff in multiple-antenna channels," *IEEE Transactions on Information Theory*, vol. 49, 2003.

[7] H. V. Trees, *Optimum Array Processing*. John Wiley & Sons, 2002.

[8] D. W. Bliss and K. W. Forsythe, "Information theoretic comparison of MIMO wireless communication receivers in the presence of interference," *Proc. 38th Asilomar Conference on Signals, Systems and Computers*, 2004.

[9] D. W. Bliss, K. W. Forsythe, I. Alfred O. Hero, and A. F. Yegulalp, "Environmental issues for MIMO capacity," *IEEE Transactions on Signal Processing*, vol. 50, no. 9, Sept. 2002.

[10] R. S. Blum, "MIMO capacity with interference," *IEEE Journal on Selected Areas of Communications*, vol. 21, no. 5, pp. 793–801, June 2003.

[11] S. Govindasamy, D. W. Bliss, and D. H. Staelin, "Spectral efficiency in single-hop ad-hoc wireless networks with interference using adaptive antenna arrays," *IEEE Journal on Selected Areas of Communications*, Sept. 2007.

[12] M. F. Demirkol and M. Ingram, "Stream control in networks with interfering MIMO links," *IEEE Wireless Communications and Networking Conference*, vol. 1, pp. 343–348, Mar. 2003.

[13] B. Chen and M. J. Gans, "MIMO communications in ad-hoc networks," *IEEE Transactions Signal Processing*, vol. 54, pp. 2773–2783, July 2006.

[14] O. B. S. Ali, C. Cardinal, and F. Gagnon, "Performance of optimum combining in a poisson field of interferers and rayleigh fading channels," *Submitted to IEEE Trans. Wireless Comm.*, 2010.

[15] R. H. Y. Louie, M. R. McKay, N. Jindal, and I. B. Collings, "Spatial multiplexing with MMSE receivers: Single-stream optimality in ad hoc networks," *To appear in Proc. IEEE Globecomm*, 2010.

[16] N. Jindal, J. G. Andrews, and S. Weber, "Rethinking mimo for wireless networks: Linear throughput increases with multiple receive antennas," in *Proc., IEEE Intl. Conf. on Communications, Dresden, Germany*, 2009.

[17] A. M. Hunter, J. G. Andrews, and S. Weber, "Transmission capacity of ad hoc networks with spatial diversity," *IEEE Transactions on Wireless Communications*, 2009.

[18] A. Hunter and J. Andrews, "Adaptive rate control over multiple spatial channels in ad hoc networks," in *Modeling and Optimization in Mobile, Ad Hoc, and Wireless Networks and Workshops, 2008. WiOPT 2008. 6th International Symposium on*, 1-3 2008, pp. 469 –474.

[19] R. Vaze and R. Heath, "Transmission capacity of ad-hoc networks with multiple antennas using transmit stream adaptation and interference cancelation," *Conference Record of the 43rd Asilomar Conference on Signals, Systems and Computers*, 2009.

[20] J. Ma, Y. J. Zhang, X. Su, and Y. Yao, "On Capacity of Wireless Ad Hoc Networks with MIMO MMSE Receivers," *IEEE Trans. Wireless Comm.*, vol. 7, no. 12, Dec. 2008.

[21] S. Hanly and D. Tse, "Resource pooling and effective bandwidths in CDMA systems with multiuser receivers and spatial diversity," *IEEE Transactions on Information Theory*, vol. 47, no. 4, pp. 1328–1351, May 2001.

[22] D. Tse and O. Zeitouni, "Linear multiuser receivers in random environments," *IEEE Transactions on Information Theory*, vol. 46, pp. 171–188, 2000.

[23] Z. Bai and J. W. Silverstein, "On the signal-to-interference-ratio of CDMA systems in wireless communications," *Annals of Applied Probability*, vol. 17, no. 1, pp. 81–101, 2007.

[24] G. M. Pan, M. H. Guo, and W. Zhou, "Asymptotic distributions of the signal-to-interference ratios of LMMSE detection in multiuser communications," *Ann. Appl. Probab.*, vol. 17, 2007.

[25] R. Couillet, M. Debbah, and J. W. Silverstein, "A deterministic equivalent for the capacity analysis of correlated multi-user mimo channels," *submitted to IEEE Trans. Inform. Theory*, vol. abs/0906.3667, 2009.

[26] F.R.Farrokhi, G. Foschini, A. Lozano, and R. Valenzuela, "Link-optimal space-time processing with multiple transmit and receive antennas," *IEEE Comm. Letters*, vol. 5, no. 3, pp. 85–87, 2001.

[27] A. F. Karr, *Probability*. Springer-Verlag, 1993.

[28] R. Horn and C. R. Johnson, *Matrix Analysis*. Cambridge University Press, 1990.

[29] N. R. Goodman, "Statistical analysis based on a certain multivariate complex gaussian distribution (an introduction)," *The Annals of Mathematical Statistics*, vol. 34, pp. 152–177, 1963.

[30] R. J. Muirhead, *Aspects of Multivariate Statistical Theory*. Wiley, 1982.

[31] T. Marzetta and B. Hochwald, "Capacity of a mobile multiple-antenna communication link in Rayleigh flat fading," *Information Theory, IEEE Transactions on*, vol. 45, no. 1, pp. 139 –157, jan 1999.

[32] A. Tulino and S. Verdu, *Random Matrix Theory and Wireless Communications*. Now Publishers, 2004.

[33] M. Haenggi, J. G. Andrews, F. Bacelli, O. Dousse, and M. Franceschetti, "Stochastic geometry and random graphs for the analysis and design of wireless networks," *IEEE J. on Selected Areas of Communications*, 2009.

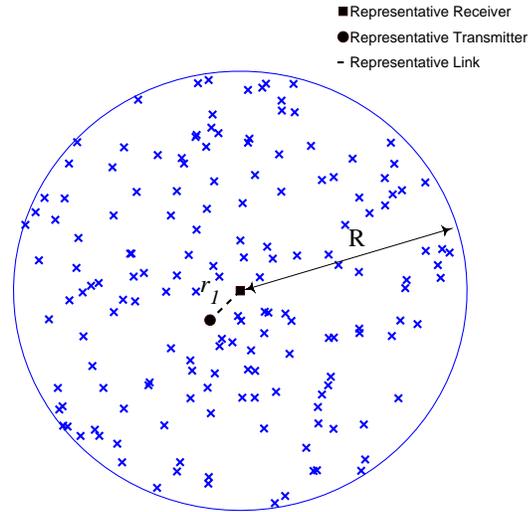

Fig. 1. Illustration of wireless network with representative link.

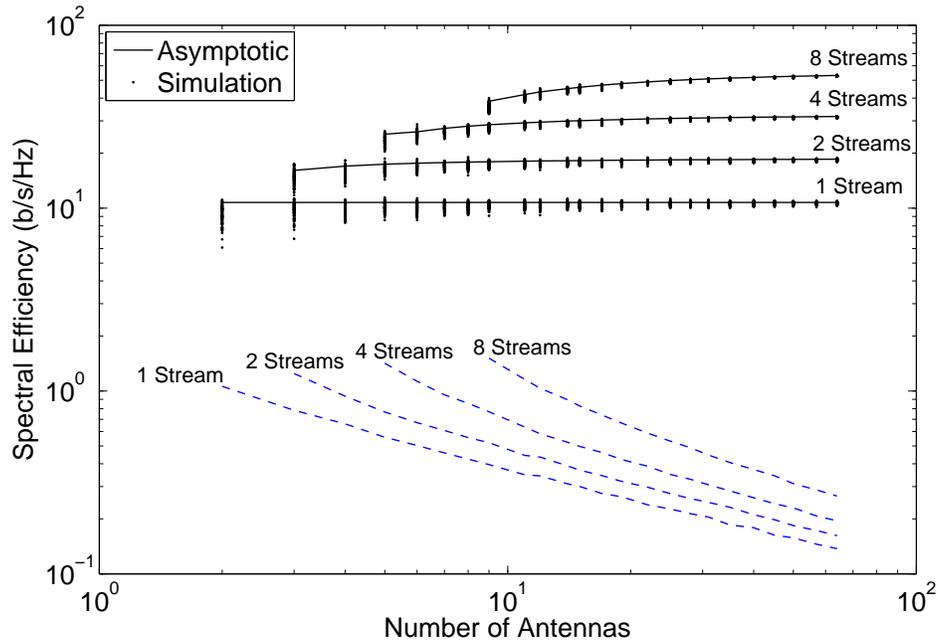

Fig. 2. Simulated and asymptotic spectral efficiency vs. number of antennas with constant path-loss, constant transmitter powers, and $n/N = 1$. The dashed lines represent the standard deviation of the simulated spectral efficiencies.

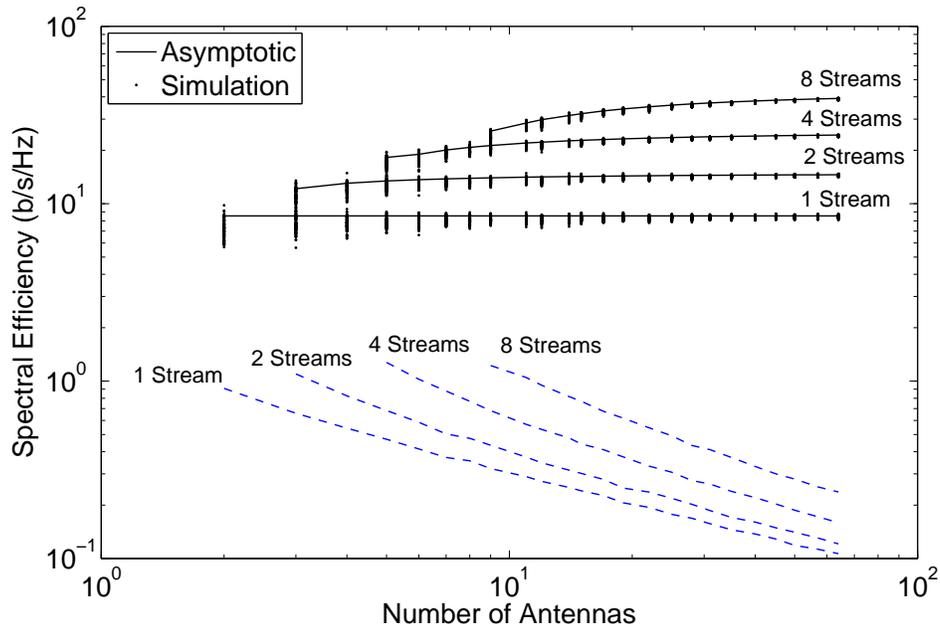

Fig. 3. Simulated and asymptotic spectral efficiency vs. number of antennas with constant path-loss, constant transmitter powers, and $n/N = 4$. The dashed lines represent the standard deviation of the simulated spectral efficiencies.

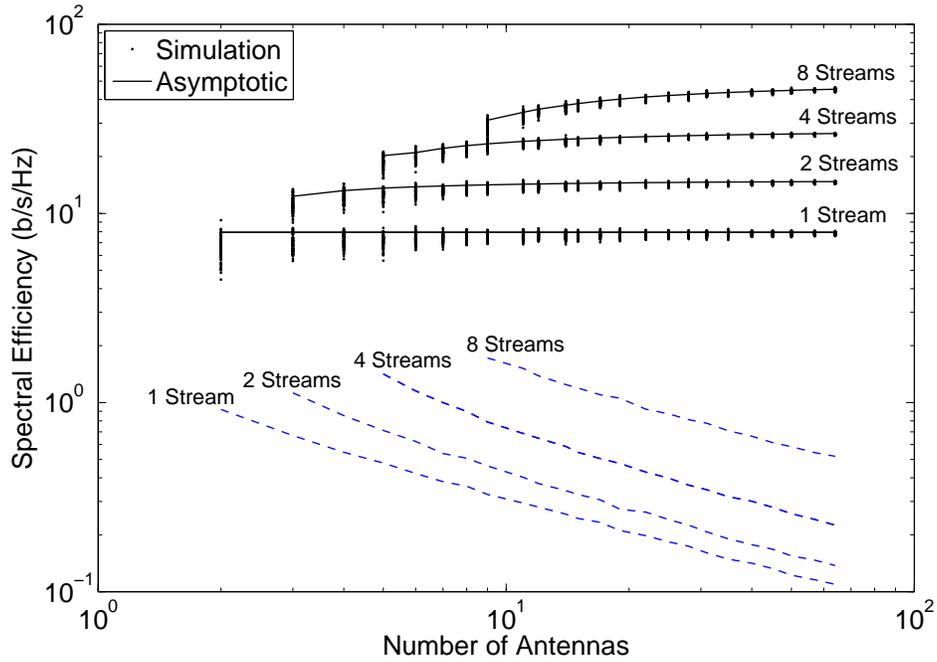

Fig. 4. Simulated and asymptotic spectral efficiency vs. number of antennas with constant path-loss, transmitter powers from the two-class model, and $n/N = 4$. The dashed lines represent the standard deviation of the simulated spectral efficiencies.

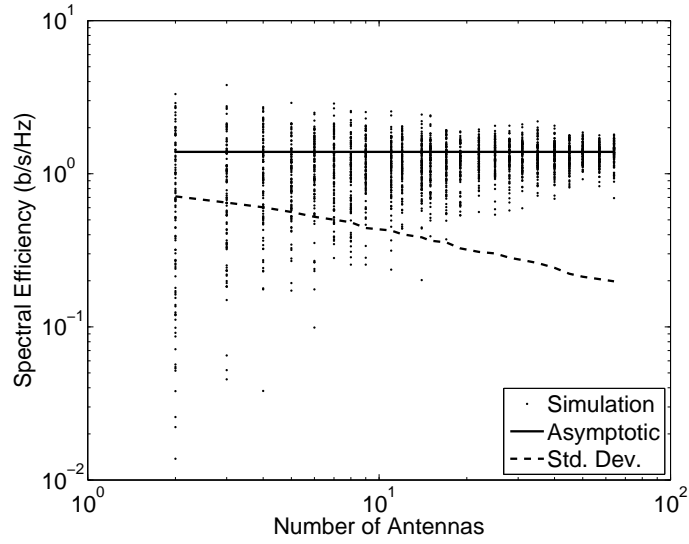

Fig. 5. Simulated and asymptotic normalized spectral efficiency vs. number of antennas for spatially distributed networks with constant transmit powers and one stream with $\alpha = 4$.

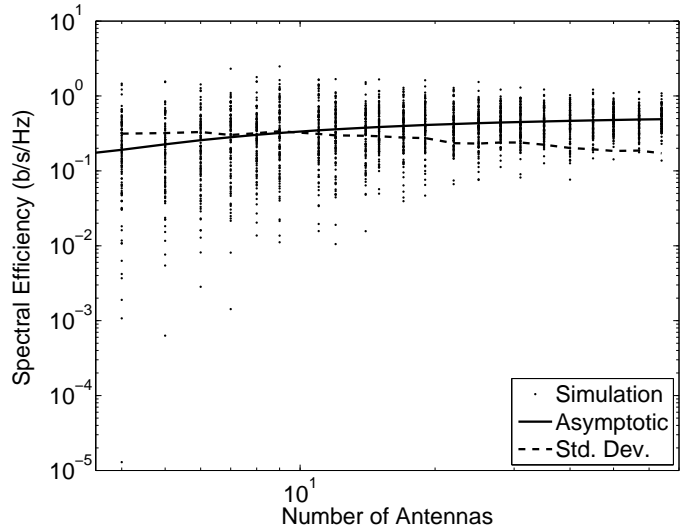

Fig. 6. Simulated and asymptotic normalized spectral efficiency vs. number of antennas for spatially distributed networks with constant transmit powers and 4 streams with $\alpha = 4$.

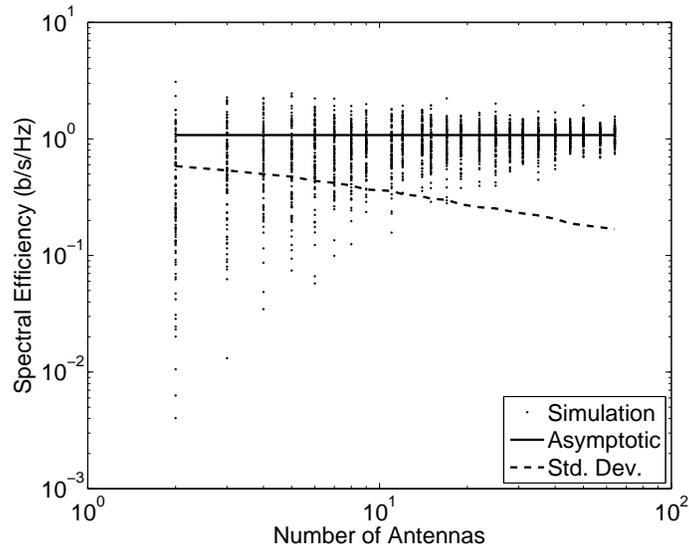

Fig. 7. Simulated and asymptotic normalized spectral efficiency vs. number of antennas for spatially distributed networks with the two-class model and one transmit stream with $\alpha = 4$.

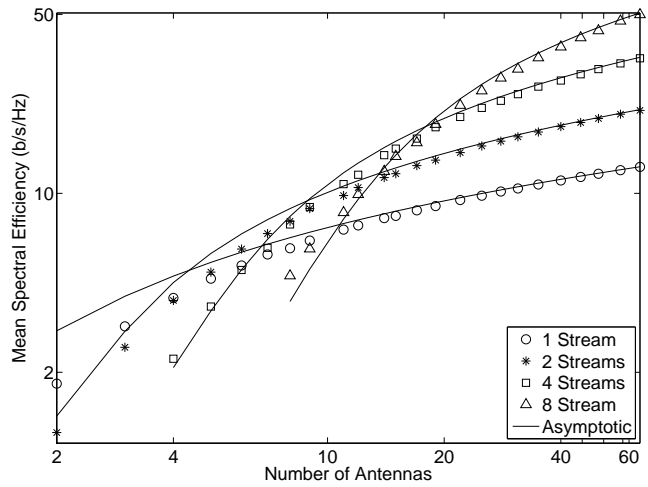

Fig. 8. Simulated and asymptotic mean spectral efficiency vs. number of antennas for spatially distributed networks with constant transmit powers and $\alpha = 4$ .

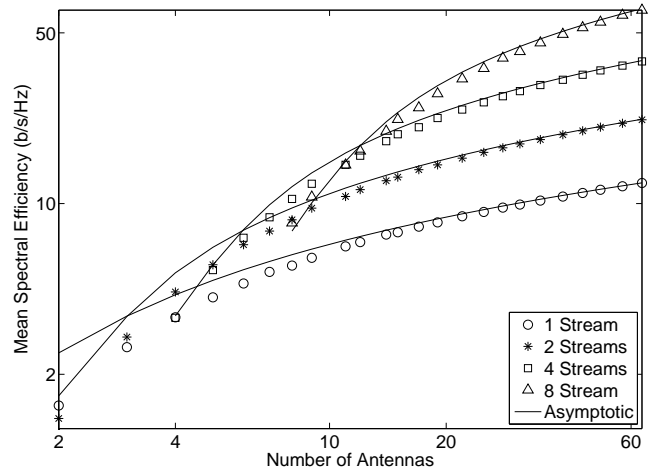

Fig. 9.  Asymptotic and simulated mean spectral efficiency vs. number of antennas for spatially distributed networks with the 2-class model and $\alpha = 4$ .

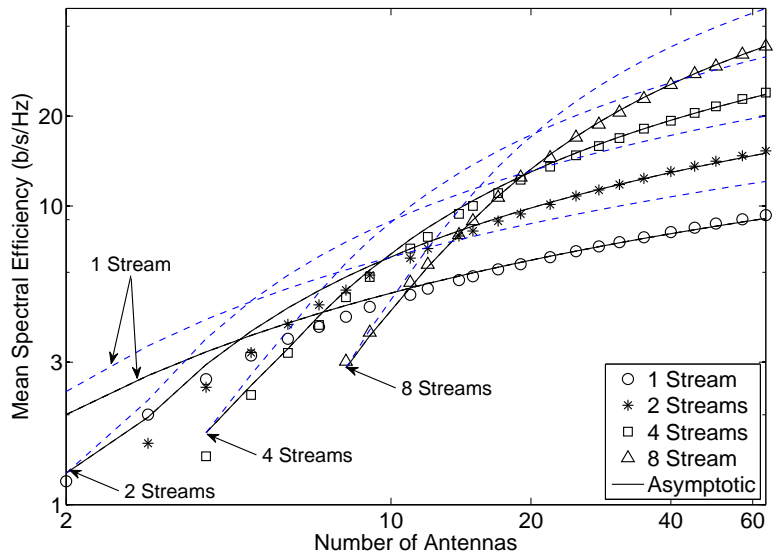

Fig. 10.  Simulated and asymptotic mean spectral efficiency vs. number of antennas for spatially distributed networks with constant transmit powers and $\alpha = 3$. The asymptotic spectral efficiency for $\alpha = 4$ is shown by the dashed lines.

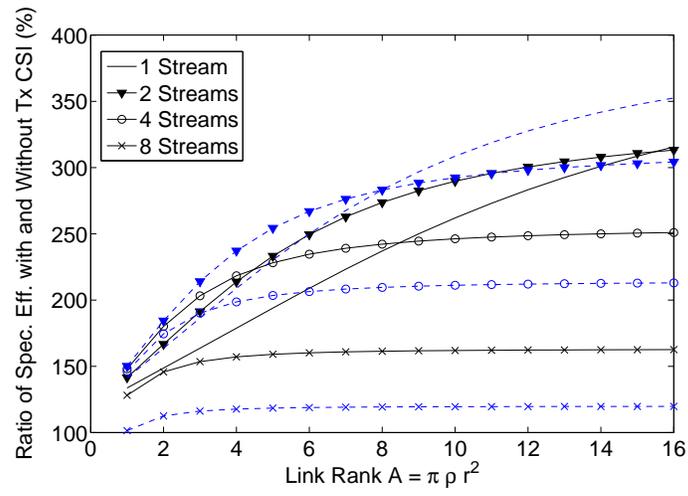

Fig. 11. Percentage increase in asymptotic mean spectral efficiency of systems with Tx Link CSI over systems without Tx CSI vs. $\pi \rho r_1^2$ for $N = 12$ (solid lines) and $N = 8$ (dashed lines) antennas per node.